\documentclass[prd,aps,superscriptaddress,showpacs,preprintnumbers,amsmath,amssymb,floatfix,12pt]{revtex4}

\usepackage{graphicx}
\usepackage{feynmp}
\preprint{CU-TP-1145}
\begin{document}
\bibliographystyle{apsrev}

\title{Relativistic Heavy Quark Effective Action}

\author{Norman H.~Christ, Min Li and Huey-Wen Lin}
\affiliation{Physics Department,Columbia University,New York, NY 10027}

\date{August 7, 2006}

\bibliographystyle{apsrev}

% ----------------------------------------------------------------
\begin{abstract}
We study the fermion action needed to accurately describe the low
energy physics of systems including heavy quarks in lattice QCD
even when the heavy fermion mass $m$ is on the order of, or larger
than, the inverse lattice spacing: $m \ge 1/a$.  We carry out an
expansion through first order in $|\vec p| a$ (where $\vec p$ is the
heavy quark momentum) and all orders in $ma$, refining the analysis
of the Fermilab and Tsukuba groups. We demonstrate that the spectrum
of heavy quark bound states can be determined accurately through
$|\vec p| a$ and $(ma)^n$ for arbitrary exponent $n$ by using a
lattice action containing only three unknown coefficients: $m_0$,
$\zeta$ and $c_P$ (a generalization of $c_{SW}$), which are functions
of $ma$.  In a companion paper, we show how these three coefficients
can be precisely determined using non-perturbative techniques.
\end{abstract}

\maketitle
\newpage

The mass spectrum and decay properties of hadrons containing charm
and bottom quarks provide some of the most precise information
about the masses and weak-interaction couplings of the underlying
quarks. In many cases, the methods of lattice QCD potentially
allow the most accurate connection between these observed hadronic
properties and the important, underlying Standard Model
parameters.  However, the relatively large masses of the charm and
bottom quarks make it difficult to perform conventional lattice
calculations since practical limitations often do not permit the
use of a sufficiently small lattice spacing to properly control
discretization errors of $O(ma)^n$.

Fortunately, in many cases those phenomena which occur at energies
on the order of the heavy quark mass are ``irrelevant'' to the
masses and matrix elements of interest, contributing only
renormalization effects which can be accounted for by the proper
choice of a small number of parameters.  In the case of a lattice
QCD calculation, this may mean that even the distortions implied
by $ma \ge 1$ can be completely compensated by such a choice of a
few parameters.

This philosophy underlies the static and non-relativistic approaches
to lattice calculations of the properties of hadrons containing heavy
quarks.  In this paper we will study a more general approach to the
treatment of heavy quarks introduced by the Fermilab
group \cite{El-Khadra:1997mp} and refined and studied in detail by
the group at Tsukuba\cite{Aoki:2001ra}.  (For a recent review of these
lattice QCD approaches to heavy quarks see Ref.~\cite{Kronfeld:2003sd}.)

In the Fermilab approach, one studies the properties of hadrons
containing heavy quarks in their rest system and argues that the
spatial momenta $\vec p$ carried by the heavy quark(s) will be
significantly smaller than the heavy quark mass:
$|\vec p| \approx \Lambda_{\rm QCD}$ for heavy-light systems and
$|\vec p| \approx \alpha_s m$ for heavy-heavy system.  Here $\alpha_s$
is the strong coupling constant evaluated at the energy scale
appropriate for the heavy-heavy bound state.  Because of the
potentially large heavy quark mass, the temporal momentum may be
very large and terms of all orders in $m a$ must be properly included.
Thus, this approximation scheme naturally treats time and space
differently, breaking the axis interchange symmetry of the usual
lattice formulation.

As will be reviewed below, accurate masses for such states
containing heavy quarks can be obtained by using a variant of the
usual Wilson action in which axis interchange symmetry is broken
and particular bare lattice parameters are chosen to be specific
functions of $ma$.  In contrast to previous work, we demonstrate
that all errors of order $|\vec p| a$ and all orders in
$ma$ may be removed by the choice of only three parameters in the
lattice action, the ratio $\zeta$ of the coefficients of the
spatial and temporal derivatives, the coefficient $c_P$ of the
axis-interchange-symmetric Pauli term and the bare fermion mass
$m_0$.

If other on-shell quantities such as matrix elements are to be
computed accurately to this order, one must also explicitly add
$ma$-dependent improvement terms to the operator whose matrix
element is being evaluated and to any interpolating field being
used to create or destroy fermion states.  For spin-1/2 states
such improvement of the interpolating field will involve both an
overall normalization factor $Z_q$ and a spinor transformation
containing a further axis-exchange-asymmetric term with an
additional mass-dependent coefficient.

Thus, apart from additional improvement coefficients needed to
evaluate specific on-shell matrix elements, only three parameters
need to be determined to carry out such $O(|\vec p| a)$,
$O(ma)^n$ lattice calculations.  As is demonstrated in
the companion paper~\cite{Lin:2006}, these three parameters can be
accurately determined from a finite-volume step-scaling procedure,
suggesting that this approach to the lattice calculation of the
properties of hadrons containing heavy fermions can be done with
no reliance on lattice perturbation theory and good control of all
systematic errors.

In the next section, Sec.~\ref{sec:analysis}, we describe this
approach to heavy quark physics in greater detail, specifying
the lattice and effective continuum actions and the field
transformations that can be used to simplify the latter.  Then
in Sec.~\ref{sec:example} we discuss a simplified example showing
why only three parameters need to be tuned in the lattice action
to achieve an accurate continuum result.  Section~\ref{sec:on_shell}
makes an explicit comparison with the results of the Fermilab and
Tsukuba groups and analyzes the disparity between the number of
parameters introduced in those treatments (four and five respectively).

Section~\ref{sec:induction} contains a complete inductive proof,
that finite lattice spacing errors of order $O(|\vec p| a)$
and $O(ma)^n$ can be removed by the proper choice of the three lattice
parameters $m_0$, $\zeta$ and $c_P$.  In Section~\ref{sec:tree} we
examine the physical heavy quark mass, on-shell quark propagator and
quark-gluon vertex at tree level to demonstrate explicitly that
$O(|\vec p| a)$ and $O(ma)^n$ accuracy requires the choice of
only three parameters and the use of an improved quark interpolating
field.  Some conclusions are presented in Sec.~\ref{sec:conclusion}.

\section{General discussion}
\label{sec:analysis}

Our objective is to describe hadrons containing heavy quarks using
a lattice action which includes a number of improvement terms
chosen to remove all finite lattice spacing errors to a given
order in $|\vec p| a$ and all orders in $ma$.  Further, as
discussed earlier, we expect that a different treatment will be
required for the spatial and temporal momenta of the heavy quark,
implying a lattice action which is axis-exchange asymmetric.  As we
will demonstrate, the desired $O(|\vec p| a)$ and $O(ma)^n$
accuracy can be achieved if we begin with a lattice fermion action
of the form
\begin{eqnarray}
S_{\rm lat} &=& \sum_{n',n} \overline{\psi}_{n'} \Bigl(
        \gamma^0 D^0 + \zeta \vec{\gamma} \cdot \vec{D}
      +  m_0 - \frac{r_t}{2} (D^0)^2 - \frac{r_s}{2} \vec{D}^2 \nonumber \\
     && + \sum_{i,j} \frac{i}{4} c_B \sigma_{ij}F_{ij}
        + \sum_{i}   \frac{i}{2} c_E \sigma_{0i}F_{0i}
       \Bigr)_{n',n}\psi_n
\label{eq:action_lat}
\end{eqnarray}
and a simple choice of the bare lattice parameters: $r_s=r_t=1$,
$c_E=c_B$.  Here $\psi_n$ is the heavy quark field at the site $n$,
$U_\mu(n)$ is the $SU(3)$ matrix providing gauge parallel transport
from the site $n+\mu$ to the site $n$ and
\begin{eqnarray}
(D_{\mu}\psi)_n & = & \frac{1}{2}\left[U_\mu(n)\psi_{n+\hat{\mu}} -
U_\mu(n-\hat{\mu})^{\dagger}\psi_{n-\hat{\mu}}\right]\\
(D_{\mu}^2\psi)_n& = &\left[U_\mu(n)\psi_{n+\hat{\mu}}
+U_\mu(n-\hat{\mu})^{\dagger}\psi_{n-\hat{\mu}}-2\psi_n\right]\\
(F_{\mu\nu}\psi)_n&=& \frac{1}{8}\!\!\!\!\sum_{s,s^\prime=\pm 1}\!\!\!\!
ss'\left[U_{s\mu}(n)U_{s^\prime \nu}(n+s\hat\mu)\right.\nonumber\\
&& \times \left.U_{-s\mu}(n+s\hat\mu+s^\prime \hat\nu)U_{-s^\prime \nu}(n+s^\prime \hat\nu)
-\mathrm{h.c.}\right]\psi_n.
\end{eqnarray}
We are using Hermitian gamma matrices $\gamma_\mu$ obeying
$\{\gamma_\mu,\gamma_\nu\}=2\delta_{\mu\nu}$ with $\sigma_{\mu\nu}
= \frac{i}{2}[\gamma_\mu,\gamma_\nu]$ and have defined the
Yang-Mills field strength tensor $F_{\mu\nu}$ to be an
anti-Hermitian color matrix.

\subsection{Continuum effective action}

In the limit that the lattice spacing becomes small we can analyze the
resulting theory and enumerate the largest lattice spacing errors by
constructing the Symanzik effective action describing a continuum
theory which approximates the lattice theory, including the discretization
errors through a given order.  Including terms representing errors of
order $a$ this effective action can be written:
\begin{eqnarray}
S_{\rm eff} &=& \int d^4 x \; \overline{\psi}(x) \Bigl(
       \gamma^0 D^0 + \zeta^c \vec{\gamma} \cdot \vec{D}
      +  m_r - a \frac{r_t^c}{2} (D^0)^2 - a \frac{r_s^c}{2} \vec{D}^2 \nonumber \\
     && + \sum_{i,j} \frac{i}{4} c_B^c    a\sigma_{ij}F_{ij}
        + \sum_{i}  \frac{i}{2} c_E^c    a\sigma_{i0}F_{i0}
        + \sum_{i}   \frac{1}{8} \delta^c a\sigma_{i0}\{D^i,D^0\}.
       \Bigr)\psi(x).
\label{eq:action_cont}
\end{eqnarray}
The superscript label $c$ representing ``continuum'' has been added to
the parameters appearing in this effective continuum action to
distinguish them from the similar parameters which enter the lattice
action of Eq.~\ref{eq:action_lat}.  The corresponding continuum mass
has been written $m_r$.

Here we are anticipating a choice of lattice parameters which
violate axis-interchange symmetry and have therefore introduced
all possible dimension 3, 4 and 5 terms which obey only the
requirement of rotational symmetry.  In Eq.~\ref{eq:action_cont}
$\psi(x)$ and $\overline{\psi}(x)$ are the usual continuum fermion
fields with normalization chosen to make the coefficient of the
$\gamma^0 D^0$ equal to $1$.  The derivatives $\vec{D}$ and
$D^0$ are the usual gauge-covariant continuum derivatives and
$F_{\mu,\nu} = [D_\mu, D_\nu]$ is the Yang-Mills field strength
tensor.

Since we are interested in treating the case where terms of the
form $(m_0a)^n$ or $(D^0 a)^n$ may be large, we will generalize
Eq.~\ref{eq:action_cont} to include correction terms containing
arbitrary powers of these two quantities but, unless accompanied
by such a factor of the heavy quark energy or mass, we neglect all other
terms of order $a^2$ or higher.  The resulting general heavy quark
Symanzik effective action might be written:

\begin{equation}
{\cal L}_{\rm eff} = {\cal L}_{\rm eff,-1}
              + {\cal L}_{\rm eff,0} + {\cal L}_{\rm eff,1} + \ldots.
                                                     \label{eq:eff}
\end{equation}
where
\begin{eqnarray}
%--------------------------------------------------
{\cal L}_{\rm eff,-1}
  &=& \overline{\psi}\Bigl( \frac{1}{a} B^{-1,1}
      + \gamma^0 D^0 C^{-1,1}\Bigr)\psi  \label{eq:eff_-1} \\
%--------------------------------------------------
{\cal L}_{\rm eff,0}
  &=& \overline{\psi}\Bigl(\{\vec\gamma \vec D, B^{0,1}\}
            + a\{[\vec\gamma \vec D, \gamma^0 D^0], C^{0,1}\}\Bigr)\psi
\label{eq:eff_0} \\
%--------------------------------------------------
{\cal L}_{\rm eff,1}
  &=& a \overline{\psi}\Bigl(\vec D^2 B^{1,1} + a\{\vec D^2, \gamma^0 D^0\} C^{1,1}
                                                                 \label{eq:eff_1} \\
  &&+ [\gamma^i,\gamma^j] [D^i,D^j] B^{1,2}
         + a\{[\gamma^i,\gamma^j] [D^i,D^j], \gamma^0 D^0 \}C^{1,2}
                                                                 \nonumber \\
  &&+ [\gamma^i,\gamma^0] [D^i,D^0] B^{1,3}
         + a[[\gamma^i,\gamma^0] [D^i,D^0], \gamma^0 D^0] C^{1,3} \Bigr)\psi.
\nonumber
\end{eqnarray}
Here the coefficient functions $B^{i,j}$ and $C^{i,j}$ are actually polynomials
of arbitrary order in the product $m_0a$, the operator $(aD^0)^2$ and the gauge
coupling $g^2$:
\begin{eqnarray}
B^{i,j}= \sum_{k,l,n} b_{k,l,n}^{i,j} (m_0a)^k\Bigl((aD^0)^2\Bigr)^{l} g^{2n} \nonumber \\
C^{i,j}= \sum_{k,l,n} c_{k,l,n}^{i,j} (m_0a)^k\Bigl((aD^0)^2\Bigr)^{l} g^{2n}.
\label{eq:coef_series}
\end{eqnarray}

Because we will work to arbitrary order in $m_0a$ and $a D^0$ it is
natural to adopt an expansion in lattice spacing $a$ where we
count only powers of $a$ which are not compensated by added powers
of $m_0$ or $D^0$.  We will refer to such an expansion as
``relativistic heavy quark'' or RHQ power counting.  The subscripts
appearing on the three terms in Eq.~\ref{eq:eff} refer to such a
scheme.

\subsection{Discrete symmetries}

The coefficients in the Symanzik effective action appearing in
Eqs.~\ref{eq:action_cont} and \ref{eq:eff_1} can be constrained
if, as is conventional, we work with an underlying lattice action
which obeys various discrete symmetries and reality conditions.
The simplest are the four symmetries corresponding to the change
in sign of one of four Euclidean coordinates:
$x_\mu \rightarrow (-1)^{\delta_{\mu\nu}}x_\mu$ where $\nu$ is the
direction being inverted.  In lattice coordinates we replace
the fields at the site with coordinates $n_\mu$ with those at the
site $n^{P(\nu)}_\mu$ where $n^{P(\nu)}_\mu = L_\mu-1-n_\mu$ for
$\mu=\nu$ and $n^{P(\nu)}_\mu = n_\mu$ otherwise.  Here we are assuming
a general space-time volume of size $L_0 \times L_1 \times L_2
\times L_3$ with $0 \le n_\mu < L_\mu$ and $L_\mu$ even.  Our lattice
fields transform as:
\begin{eqnarray}
\psi_n            &\rightarrow& \gamma^\nu\gamma^5\psi_{n^{P(\nu)}} \\
\overline{\psi}_n &\rightarrow& \overline{\psi}_{n^{P(\nu)}}\gamma^5\gamma^\nu    \\
U_\nu(n)          &\rightarrow&
            U_\nu^\dagger\Bigl(n^{P(\nu)}-\hat{e_\nu}(1- L \delta_{n_\nu,L_\nu-1})\Bigr) \\
U_\mu(n)          &\rightarrow& U_\mu(n^{P(\nu)}) \quad \mbox{for $\mu \ne \nu$}.
\end{eqnarray}
Here $\hat e_\nu$ is a vector extending one site in the $\nu$ direction.

For $ma \ge 1$ the mass shell condition $p_0 = +m$ will imply that
the negative energy, anti-quark states are far outside the domain of
validity of our approximation.  Thus, it is important that the improved
lattice action obey charge-conjugation symmetry so that both heavy quarks
and heavy anti-quarks will be treated with the same accuracy.  This can
be accomplished if we require that our improved lattice action and therefore
the continuum Symanzik action are symmetric under the following change
of variables:
\begin{eqnarray}
\psi_n               &\rightarrow& C\overline{\psi}_n^t \label{eq:cc1}\\
\overline{\psi}_n    &\rightarrow& -\psi_n^tC^{-1}      \label{eq:cc2}\\
U_\mu(n)             &\rightarrow& U_\mu(n)^*           \label{eq:cc3}
\end{eqnarray}
where the Dirac charge conjugation matrix $C$ obeys
\begin{equation}
C^{-1}\gamma_{\mu}C = -\gamma_{\mu}^t.
\end{equation}
Here we are treating the Grassmann variables $\psi$ and $\overline{\psi}$
as $4 \times 1$ and $1 \times 4$ spinor matrices respectively which
requires the appearance of the transpose operation in Eqs.~\ref{eq:cc1}
and \ref{eq:cc2} indicated by the superscript $t$.

The lattice action given in Eq.~\ref{eq:action_lat} already obeys the
above axis reversal symmetry given our requirement that only even powers
of the operator $aD_0$ appear.  All of the terms in Eq.~\ref{eq:action_lat}
are also charge conjugation even except for the terms containing the
functions $C^{0,1}$ and $C^{1,3}$.  These are odd under $C$ and can
be set to zero.

Finally we should determine the phases of the coefficients appearing
in the effective action of Eq.~\ref{eq:eff}.  We begin with the lattice
action given in Eq.~\ref{eq:action_lat}.  Here we have introduced
factors of $i$ in such a way that this action will yield a
Hermitian~\cite{Luscher:1976ms}, but possibly not
positive~\cite{El-Khadra:1997mp} transfer matrix if the bare, lattice
parameters $m_0$, $\zeta$, $r_s$, $r_t$, $c_B$ and $c_E$ are all chosen
real.

The phases of the parameters appearing in the continuum effective
action of Eq.~\ref{eq:eff} can then be easily constrained if we recognize
that when the above 6 bare lattice parameters are real, the underlying
lattice action obeys a simple symmetry under complex conjugation.
Specifically, we consider the fermion path integral in a fixed gauge
background:
\begin{equation}
Z[\eta,\overline{\eta}] = \int d[\psi] d[\overline\psi]
       \exp\Biggl\{S[\psi,\overline\psi]_{\rm lat}
                   +\int d^4x \bigl\{ \overline\psi(x)\eta(x)
                   + \overline\eta(x)\psi(x) \bigr\}\Biggr\},
\label{eq:gen_fctn}
\end{equation}
where we have introduced explicit sources $\eta$ and
$\overline{\eta}$ so that arbitrary Green's functions can be
determined.  The integral in Eq.~\ref{eq:gen_fctn} will evaluate
to a polynomial in the Grassmann variables $\eta$ and
$\overline{\eta}$ with complex coefficients. If we define
$Z[\eta,\overline{\eta}]^*$ as that same polynomial but with the
coefficients replaced by their complex conjugates, then one can
easily show by a standard change of variables in the path integral
in Eq.~\ref{eq:gen_fctn},
\begin{equation}
\psi          \rightarrow \gamma^5\overline\psi^t \quad
\overline\psi \rightarrow -\psi^t\gamma^5,
\label{eq:ccc_trans}
\end{equation}
that when $m_0$, $\zeta$, $r_s$, $r_t$, $c_B$ and $c_E$ are real the
following relation is obeyed:
\begin{equation}
Z[\eta,\overline{\eta}]^* = Z[\gamma^5\overline{\eta}^t, -\eta^t\gamma^5].
\label{eq:ccc_cond}
\end{equation}

Since the continuum effective action is determined directly from
the lattice action, it also must obey this reality condition.
This requires that each of the functions $B^{i,j}$ and $C^{i,j}$
appearing in Eq.~\ref{eq:eff} be polynomials in the three
quantities $m_0$, $(aD^0)^2$ and $g^2$ with real coefficients.

\subsection{Field transformations}

As is well known, many of the unwanted terms in Eq.~\ref{eq:eff} have
no effect on physical states or fermion Green's functions evaluated on
the mass shell and can be removed by a redefinition of the fermion fields
$\psi$ and $\overline{\psi}$.  We will therefore make a series of such
transformations chosen to remove many of the terms that appear in the
Symanzik effective action of Eq.~\ref{eq:eff}.  The coefficients
of those terms that remain after these transformations are then presumed
to be potentially important lattice artifacts that must be eliminated
by an explicit choice of additional improvement terms in the underlying
lattice action.

The removal of these redundant terms is most easily analyzed in a series
of steps exploiting the ordering of the terms in Eq.~\ref{eq:eff}:
$O(1/a)$, $O(a^0)$, $O(a)$, {\it etc.} in the RHQ expansion.  The largest
field transformation introduces terms of order $a^0$ in this RHQ expansion
and can be written:
\begin{eqnarray}
\psi            &=& (1+R^{0,1} + a\gamma^0 D^0 S^{0,1})\psi'
\label{eq:field_trans_0a} \\
\overline{\psi} &=& \overline{\psi}'(1+\overline{R}^{0,1}
                     - a\gamma^0 \overleftarrow{D}^0 \overline{S}^{0,1}),
\label{eq:field_trans_0b}
\end{eqnarray}
where $R^{0,1}$, $S^{0,1}$, $\overline{R}^{0,1}$ and $\overline{S}^{0,1}$
are arbitrary polynomials in $m_0a$, $(aD^0)^2$ and $g^2$.  We adopt the
convention in the transformation equations above and the four to follow,
that the $aD^0$ argument will always act to the right in the equations
for $\psi$ and to the left in the equations for $\overline{\psi}$.  (Note
that as the covariant derivative, the operator $D_\mu$ will have a
different form when acting on $\overline{\psi}$, a color vector whose
gauge transformation properties are the hermitian conjugate of those of $\psi$,
see Appendix A, Eqs.~\ref{eq:cov_derivative_1} and \ref{eq:cov_derivative_2}.)
This transformation will effect all three terms shown in Eq.~\ref{eq:eff},
$O(a^{-1})$,  $O(a^{0})$ and $O(a^{1})$ and will generate extra terms that
can be used to simplify the resulting action.

Relevant to the order in $a$ to which we are working are two further
transformations.  The first transformation introduces terms of order $a^1$
in $\psi$ and $\overline{\psi}$ and takes the form:
\begin{eqnarray}
\psi            &=& (1+ a \vec\gamma \vec D R^{1,1}
                 + a[\vec\gamma \vec D,a\gamma^0 D^0] S^{1,1})\psi'
\label{eq:field_trans_1a} \\
\overline{\psi} &=& \overline{\psi}'(1
                - a \overline{R}^{1,1} \vec\gamma \overleftarrow{D}
                - a \overline{S}^{1,1} [\vec\gamma \overleftarrow{D},a\gamma^0
                           \overleftarrow{D}^0] ).
\label{eq:field_trans_1b}
\end{eqnarray}
This transformation will act on the ${\cal L}_{{\rm eff},-1}$ and
${\cal L}_{{\rm eff},0}$ terms in Eq.~\ref{eq:eff} and produce
terms of order $a^0$ and $a^1$ in the transformed action.

Finally, we must discuss a third transformation which is of order $a^2$:
\begin{eqnarray}
\psi            &=& \Biggl(1+ a^2 \vec D^2 R^{2,1}
                   + a^2\{\vec D^2,a\gamma^0 D^0\}S^{2,1}
\label{eq:field_trans_2a} \\
                 &&+a^2[\gamma^i, \gamma^j][D^i, D^j] R^{2,2}
    +a^2\Bigl\{[\gamma^i, \gamma^j][D^i, D^j],a\gamma^0 D^0\Bigr\}S^{2,2}
\nonumber \\
   &&+a^2[\gamma^i, \gamma^0][D^i, D^0] R^{2,3}
   +a^2\Bigl[[\gamma^i, \gamma^0][D^i, D^0],a\gamma^0 D^0\Bigr]S^{2,3}
\Biggr)\psi' \nonumber \\
\overline{\psi} &=& \overline{\psi}'\Biggl(1
                   + a^2 \overleftarrow{D}^2\overline{R}^{2,1}
 - a^2\{\overleftarrow{D}^2,a\gamma^0 \overleftarrow{D}^0\}\overline{S}^{2,1}
\label{eq:field_trans_2b}
 \\
        &&+a^2[\gamma^i, \gamma^j][\overleftarrow{D}^i, \overleftarrow{D}^j]
                            \overline{R}^{2,2}
          -a^2\Bigl\{[\gamma^i, \gamma^j][\overleftarrow{D}^i, \overleftarrow{D}^j],
               a\gamma^0 \overleftarrow{D}^0\Bigr\}\overline{S}^{2,2}
\nonumber \\
        &&+a^2[\gamma^i, \gamma^0][\overleftarrow{D}^i, \overleftarrow{D}^0]
                 \overline{R}^{2,3}
          +a^2\Bigl[ [\gamma^i, \gamma^0][\overleftarrow{D}^i, \overleftarrow{D}^0],
             a\gamma^0 \overleftarrow{D}^0\Bigr]\overline{S}^{2,3} \Biggr).
\nonumber
\end{eqnarray}
This order $a^2$ transformation was not investigated in Ref.~\cite{Aoki:2001ra}
nor in Section III on redundant couplings in Ref.~\cite{El-Khadra:1997mp}
although later in that paper this transformation is discussed, see Eq.~5.23.

This transformation acts on only the ${\cal L}_{{\rm eff},-1}$
term in Eq.~\ref{eq:eff} to produce terms of order $a^1$ in the
transformed action.  The effects of these transformations will be
considered below, first in a simplified context in
Sec.~\ref{sec:example} and then in generality in
Sec.~\ref{sec:induction}.   Here we will specialize these three
transformations to preserve the charge conjugation symmetry and
reality properties discussed above.

In fact, with the choice of signs in
Eqs.~\ref{eq:field_trans_0a}-\ref{eq:field_trans_2b}, charge conjugation
requires, $R^{i,j}=\overline{R}^{i,j}$ and $S^{i,j}=\overline{S}^{i,j}$
while preservation of the form of the reality condition requires that
all coefficients in the polynomials $R^{i,j}$ and $S^{i,j}$ be real.

This completes our general discussion of the lattice action, the resulting
effective continuum action and the field transformations that can be
applied to that effective action consistent with our charge conjugation
and reality conditions.

\section{Simplified example}
\label{sec:example}

In Sec.~\ref{sec:induction} we use induction to apply the field
transformations discussed in the previous section to
systematically eliminate all terms from the effective action of
Eq.~\ref{eq:eff} except for three, mass-dependent coefficients.
These field transformations will leave the effective action in the
form given in Eq.~\ref{eq:action_cont} with only the three
coefficients $m_r$, $\zeta^c$ and $c^c_P \equiv c_B^c=c_E^c$
non-zero functions of the quark mass times the lattice spacing,
$m a$. However, in this section we will present this argument in
a simplified case which should make the conclusion and the
essential ingredients needed to reach it easier to understand.

We will consider the case that the effective continuum action is
determined by the Lagrangian given in Eq.~\ref{eq:action_cont} which
can be written:
\begin{eqnarray}
S_{\rm eff} &=& \sum_n \overline{\psi}(x) \Bigl(
        \gamma^0 D^0 + \zeta^c \vec{\gamma} \cdot \vec{D}
      +  m_r - a \frac{r_t^c}{2} (D^0)^2 - a \frac{r_s^c}{2}
              \vec{D}^2 \nonumber \\
     && + \sum_{i,j} \frac{i}{4} c_B^c    a\sigma_{ij}F_{ij}
        + \sum_{i}  \frac{i}{2} c_E^c    a\sigma_{i0}F_{i0} \Bigr)\psi(x).
\label{eq:action_cont2}
\end{eqnarray}
Here we are simplifying the general problem by dropping potentially
large time derivative terms, $(aD^0)^{2n}$, beyond those appearing
explicitly in Eq.~\ref{eq:action_cont2}.  We have also omitted the
final term proportional to $\delta$ in Eq.~\ref{eq:action_cont}
since it violates charge conjugation symmetry.

Through a combination of tuning the bare lattice parameters and
redefinition of the fields $\psi$ and $\overline{\psi}$ we will be able
to put the Lagrangian above into the standard continuum form:
\begin{equation}
{\cal L}_{\rm eff}
         = \overline{\psi^\prime}\{\gamma^0 D^0 + \gamma^i D^i + m_r \}\psi^\prime.
\label{eq:cont}
\end{equation}
As is conventional, we will work backward from Eq.~\ref{eq:cont}, performing
transformations on the fields $\psi^\prime$ and $\overline{\psi^\prime}$
in an attempt to generate as many as possible of the terms appearing
in Eq.~\ref{eq:action_cont2}.  We can then be guaranteed that if the other
terms, not created by these transformations, are set to zero by tuning an
improved lattice action, these remaining terms can then be eliminated by a
field transformation.

Let us now extend the usual transformations $\psi^\prime \rightarrow \psi$
and $\overline{\psi}^\prime \rightarrow \overline{\psi}$ to demonstrate
the redundancy of all but the three parameters listed above: $m_0a$, $\zeta$,
$c_{P}$.  As in the more complete discussion of Sec.~\ref{sec:induction},
we will organize this discussion using RHQ power counting where the
quantities $m$ and $D^0$ are treated as order $a^{-1}$ instead of $a^0$.

We begin by making transformations of $O(a^0)$ in the RHQ power counting
sense and $O(a^1)$ in the usual sense:
\begin{eqnarray}
\psi^\prime            &=& (1 + a \gamma^0 D^0 S^{0,1})\psi
\label{eq:trans_0a} \\
\overline{\psi}^\prime &=& \overline{\psi}(1
                        - a \gamma^0 \overleftarrow{D}^0 S^{0,1}),
\label{eq:trans_0b}
\end{eqnarray}
where the function $S^{0,1}$ is real and Eqs.~\ref{eq:trans_0a} and
Eqs.~\ref{eq:trans_0b} are related by charge conjugation symmetry.  We have
adopted a somewhat cumbersome notation that will be useful later: the
first integer in the superscript of $S^{i,j}$ identifies the RHQ power
counting order of the transformation and the second enumerates the different
terms of that order.  This transformation generates two terms when acting
on the action of Eq.~\ref{eq:cont}:
\begin{eqnarray}
\overline{\psi}\Bigl\{2 m_r a \gamma^0 D^0 S^{0,1} +
            2 S^{0,1} a(D^0)^2 \Bigr\}\psi.
\label{eq:action_trans_0}
\end{eqnarray}
As is customary, we neglect terms quadratic in  $S^{0,1}$ treating
these terms as small.  (This issue will be dealt with in
a more systematic way in Sec.~\ref{sec:induction}.)  Since the quantity
in Eq.~\ref{eq:action_trans_0} is generated by a change of Grassmann
variables in the path integral, we can treat such a combination of
terms as zero, were it to appear in the effective Lagrangian of our
improved lattice theory.  Of course, since by construction the
expression in Eq.~\ref{eq:action_trans_0} is linear in the Dirac
operator appearing in the final action, one can also describe the
vanishing of these terms as a consequence of the equations of motion.
These two styles of derivation are really one and the same.

The vanishing of the combination of terms in Eq.~\ref{eq:action_trans_0}
implies we can adjust the function $S^{0,1}$ to set $r_t^c$ to zero.
(Note, this gives us the freedom in the improved lattice Lagrangian to
choose the conventional value of 1 for the bare version of $r_t$.)  The
only effect on the resulting action will be that of the first term,
$2 m_r a S^{0,1} \overline{\psi}\gamma^0 D^0\psi$, which can be removed by
a rescaling of $\psi$ and $\overline{\psi}$.

Next consider transformations of order $O(a^1)$ in the RHQ power counting
sense and also $O(a^1)$ in the usual sense:
\begin{eqnarray}
\psi^\prime            &=& (1 + a \vec\gamma \vec D R^{1,1})\psi
\label{eq:trans_1a} \\
\overline{\psi}^\prime &=& \overline{\psi}(1
                        - a \vec\gamma \overleftarrow{D} R^{1,1}).
\label{eq:trans_1b}
\end{eqnarray}
Acting on the continuum Lagrangian in Eq.~\ref{eq:cont}, These
transformations will produce the terms
\begin{eqnarray}
\overline{\psi}\Bigl(2 m_r a \vec\gamma \vec D +
\frac{1}{2}a[\gamma^i,\gamma^0] [D^i,D^0]
                 + 2 a(D^i)^2 \Bigr)R^{1,1}\psi.
\label{eq:action_trans_1}
\end{eqnarray}
Hence with a proper choice for $R^{1,1}$ we can use the $a(D^i)^2$ in
Eq.~\ref{eq:action_trans_1} to set $r_s^c=0$ for any choice of $r_s$ in the
bare lattice Lagrangian (including our conventional value $r_s=\zeta$).

Thus, using the set of two transformation considered so far we
have been able to argue that an effective Lagrangian with any set
of values of $r_s$ and $r_t$ can be transformed to the proper
continuum form.  This is the standard argument reducing the number
of relevant parameters from six to four.  However, there is one
further transformation which is of $O(a^2)$ in the sense of both
RHQ and conventional power counting that can remove one more
parameter:
\begin{eqnarray}
\psi^\prime            &=& (1 + a^2[\gamma^i,\gamma^0][D^i,D^0] R^{2,3})\psi
\label{eq:trans_2a} \\
\overline{\psi}^\prime &=& \overline{\psi}(1
            + a^2[\gamma^i,\gamma^0][\overleftarrow{D}^i,\overleftarrow{D}^0]
                  R^{2,3}),
\label{eq:trans_2b}
\end{eqnarray}
where we using the label $R^{2,3}$ to maintain consistency with
Eqs.~\ref{eq:field_trans_2a} and \ref{eq:field_trans_2b}.  This
transformation, when acting on the two $O(1/a)$ terms in the continuum
action, will produce the following combination of terms of $O(a^1)$
according to RHQ power counting:
\begin{eqnarray}
a^2\overline{\psi}
      \Bigl(2m_r [\gamma^i,\gamma^0][D^i,D^0]
           +\gamma^i \Bigl[[D^i,D^0],D^0\Bigr] \Bigr) R^{2,3}\psi.
\label{eq:action_trans_2}
\end{eqnarray}
As before, we can treat this combination of terms as vanishing either
because they were generated by a transformation of path integration
variables or as a result of the equations of motion since it was obtained
as a sum of left and right multiplication by the continuum Dirac operator.

While the first term in Eq.~\ref{eq:action_trans_2} involves the usual
$\sigma^{i0} F^{i0}$ associated with $c_E$ and is nominally of order
$a$ in RHQ power counting, the second term in which both factors of $D^0$
appear in commutators has no compensating factor of $m$ and hence is $O(a^2)$.
Thus, the vanishing of the sum of terms in Eq.~\ref{eq:action_trans_2}
on-shell implies that the $c_E^c$ term in the effective action can be related
to other terms that are explicitly of order $a^2$ in the sense of RHQ
power counting.  Because of the presence of the $m_r a$ factor appearing
in this term, we cannot completely remove the $c_E$ term in the effective
action since that term will contain contributions that are not proportional
to the mass.  However, the difference between $c_E$ and $c_B$ can be
arranged to be proportional to the heavy quark mass.  We must merely avoid
a gratuitous violation of axis-interchange symmetry when choosing arbitrary
parameters in the lattice Lagrangian, {\it e.g.} we must choose
$r_t - r_s \propto (m_ra)^1$.  That is, if only axis-interchange asymmetry
proportional to $m_r a$ is introduced, the difference between $c_E$ and $c_B$
will also vanish as $m_r a \rightarrow 0$.  Thus, we can adjust the
transformation parameter $R^{2,3}$ to set $c^c_E=c^c_B \equiv c^c_{P}$.  Note,
in the limit $m_r a \ll 1$, $c_P(m_r a) \rightarrow c_{SW}$, the usual
Sheikholeslami and Wohlert coefficient of Ref.~\cite{Sheikholeslami:1985ij}.

It is natural to consider also a transformation of $O(a^2)$ in which
the coefficient of $c_B$ appears:
\begin{eqnarray}
\psi^\prime            &=& (1 + a^2[\gamma^i,\gamma^j][D^i, D^j] R^{2,2})\psi
\label{eq:trans_2c} \\
\overline{\psi}^\prime &=& \overline{\psi}(1
            + a^2[\gamma^i,\gamma^j][\overleftarrow{D}^i, \overleftarrow{D}^j]
                  R^{2,2}).
\label{eq:trans_2d}
\end{eqnarray}
However, in contrast to the previous transformation in Eqs.~\ref{eq:trans_2a}
and \ref{eq:trans_2b}, this transformation results in a collection of
terms which involves the combination: $\Bigl\{[\gamma^i,\gamma^j][D^i,D^j],
\gamma^0 D^0\Bigr\}$.  This is a new term, not included in the simplified
action of Eq.~\ref{eq:action_cont2}, which is nominally of order $a$ in our
RHQ power counted scheme and hence potentially significant.  Replacing the
$c_B$ term with this one is merely trading one non-redundant term for another.
Of course, the appearance of this new term indicates the limitations of our
simplified example and motivates the complete discussion given in
Sec.~\ref{sec:induction}.

This result that the difference between $c_B^c$ and $c_E^c$ in the
continuum effective Lagrangian contributes a term of order
$(\vec p a)^2$ can be understood qualitatively as
follows.  In the case that $m_r \ll 1/a$ we are dealing with the
standard $O(a)$ improvement of Sheikholeslami and Wohlert with
$c_B^c = c_E^c$.  To the extent that $m_r \approx 1/a$, asymmetries
between space and time will be visible and we expect $c_B^c-c_E^c
\propto m_ra$. However, for such a heavy quark case we expect the
matrix elements of the correction terms
$\overline{\psi}\sigma_{\mu\nu} F^{\mu\nu}\psi$ to be of order
$1/m_r$.  The resulting combination of an overall factor of $a$
present because this is a dimension-5 correction term, the factor
$m_r a$ coming from $c_B^c-c_E^c$ and this $1/m_r$ estimate gives an
over-all size of $O(a^2)$ with no compensating factors for $m_r$,
demonstrating that their difference can be neglected to our
intended order of accuracy.

Thus, to construct an improved lattice Lagrangian which will yield heavy
quark spectral quantities which are accurate up to but not including
$O(\vec p a)^2$ we need only tune 3 lattice parameters: $m_0$, $\zeta$,
and $c_P$.

\section{On-shell improvement and earlier work}
\label{sec:on_shell}

In the previous sections we have determined the number of
parameters that must be tuned in the lattice action if the
resulting effective continuum action is to be equivalent to
the standard continuum fermion action after a redefinition
of the fermion fields.  In this section we will consider
the limitations of the resulting improved theory and its
relation to the results of the Fermilab~\cite{El-Khadra:1997mp}
and Tsukuba~\cite{Aoki:2001ra} groups.

As is well known, the physical masses determined by an effective
theory are not changed by a change of field variables in the
path integral defining the Green's functions of that theory.
This observation underlies the reduction of parameters that we
have been investigating.  Those parameters that can be removed
by a redefinition of fields cannot effect the predicted masses.
In the earlier work of the Fermilab group, the total number of
parameters remaining after compensating for the redundancy implied
by field transformations was given as four: $m_r$, $\zeta^c$,
$c_B^c$ and $c_E^c$ in our notation.  By considering the additional
field transformation given in Eqs.~\ref{eq:trans_2a} and \ref{eq:trans_2b}
we have shown that the number of relevant parameters can be reduced
to three: $m_r$, $\zeta^c$, $c_P^c$.

Before comparing with the results of the Tsukuba group, we should
discuss the question of computing on-shell Green's functions with
the effective actions under consideration.  While our ability to
remove redundant terms from the action is established by
examining possible field transformations, such transformations
are not actually made.  Making these field transformations and
casting the effective action in the desired continuum form would
require knowing these extra, ``redundant'' parameters.  Thus, the
quark fields that appear in a lattice calculation with properly
tuned values for the three relevant input parameters (here denoted
$\psi_0$ and $\overline{\psi}_0$) are un-transformed fields which
correspond to an effective action which is not in the continuum
form.

\begin{figure}
%\vskip -1.5in
%\centering
%\begin{fmffile}{wfrnrm}
%  \begin{fmfchar*}(80,120)
%     \fmfright{qb}
%     \fmfv{dec.shape=cross,dec.filled=full,dec.size=15}{qb}
%     \fmfv{label=\large $\overline{\psi}^c$,label.dist=20}{qb}
%     \fmfpoly{hatch,tension=40,smooth}{qr,g1,g2,p0,ql,p1,p2,p3}
%     \fmfleft{q}   \fmflabel{\large $\psi^0$}{q}
%     \fmf{gluon,right=0.6,tension=0.1}{qb,g1}
%     \fmf{gluon,right=1.0,tension=0.1}{qb,g2}
%     \fmf{fermion,tension=20}{qb,qr}
%     \fmf{fermion,tension=20}{ql,q}
% \end{fmfchar*}
%\end{fmffile}
\begin{center}
\includegraphics[width=\columnwidth]{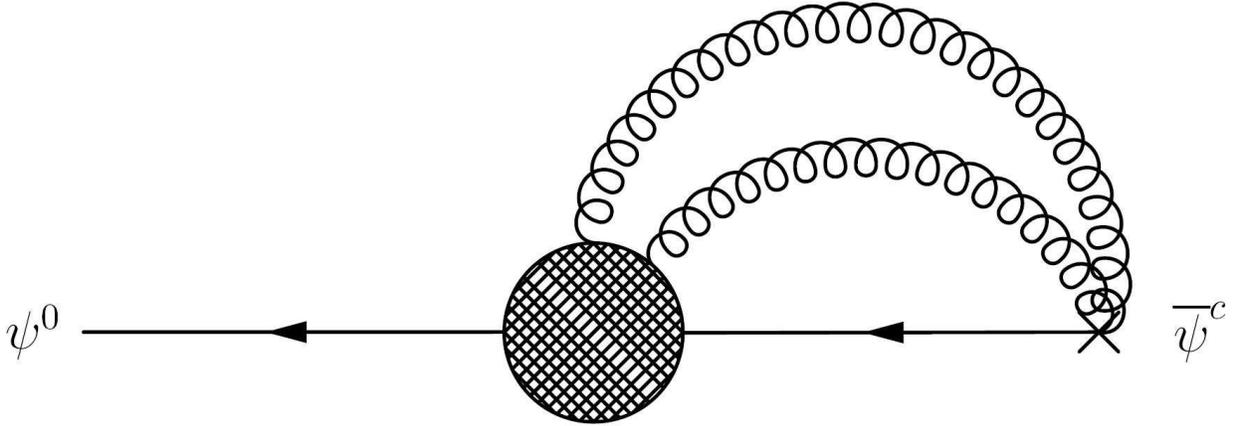}
\end{center} \caption{A class of diagrams contributing to the $4
\times 4$ spinor renormalization matrix
$\overline{Z}_{\alpha\beta}$ connecting the improved and
un-improved fields, $\overline{\psi}^c$ and $\overline{\psi}^0$
respectively.  Here the point-like vertex represented by the cross
corresponds to the composite, improved operator $\psi^c$ which
contains products of the quark and gluon fields.  The graph
contained within the shaded circle must be
one-particle-irreducible.} \label{fig:wf_renorm}
\end{figure}

Thus, we will obtain appropriate, continuum on-shell Green's
functions only after we relate the un-transformed, interpolating
fields appearing in a lattice calculation with the transformed
fields corresponding to a proper, continuum-like effective theory
(here labeled $\psi^c$ and $\overline{\psi}^c$).  While the fields
$\psi_0$,$\overline{\psi}_0$ and $\psi^c$, $\overline{\psi}^c$
are related by a complicated transformation, non-linear in the
gluon fields, we need to relate only their on-shell matrix
elements.  For such ``pole'' contributions all of the added
powers of the gluon field present in the lattice fields $\psi_0$
and $\overline{\psi}_0$ must be contracted within field
renormalization subdiagrams.  (These are one-particle-irreducible
subdiagrams with two external lines, which contain the external
quark line and the internal quark line contributing to the single
particle pole, illustrated in Fig.~\ref{fig:wf_renorm}.)  Thus,
for the purposes of evaluating on-shell Green's functions these
two sets of fields are related by a simple spinor renormalization
factor:
\begin{eqnarray}
(\psi_0)_\alpha            &=& \sum_\beta Z_{\alpha,\beta} (\psi^c)_\beta \\
(\overline{\psi}_0)_\alpha &=& \sum_\beta (\overline{\psi}^c)_\beta
                                   \overline{Z}_{\beta,\alpha}
\end{eqnarray}
Here $Z_{\alpha,\beta}$ is a simple $4 \times 4$ spinor matrix
that can be written:
\begin{equation}
Z = Z_1 + Z_2 a \vec\gamma \vec \partial.
\end{equation}
Here the coefficients $Z_i$ are arbitrary polynomials $m_0a$ and
$(\partial_0 a)^2$.  Imposing charge conjugation and reality constraints we
find that
\begin{equation}
\overline{Z} = Z_1 - Z_2 a \vec\gamma \overleftarrow{\partial}.
\end{equation}
and that the polynomials $\{Z_i\}_{i=1,2}$ have real coefficients.
Note, we have used the equations of motion to remove a possible
$\gamma^0\partial_0$ term.  Since these relations are only
to be used on-shell, the argument $(\partial^0 a)^2 =
 (m_r a)^2 + (\vec p a)^2$ and we can drop the final $(\vec p a)^2$
term.  Thus, we will adopt the form
\begin{eqnarray}
Z            &=& Z_q^{-1/2}(1+\delta a \vec \gamma \vec \partial)
\label{eq:Z_factor_a} \\
\overline{Z} &=& Z_q^{-1/2}(1-\delta a \vec \gamma \overleftarrow \partial).
\label{eq:Z_factor_b}
\end{eqnarray}
where $Z_q$ and $\delta$ are functions of $m a$ only.

Thus, our failure to actually transform to the proper continuum
fields requires that on-shell Green's functions in which the
quark fields appear as interpolating fields must have the
additional renormalization matrices $Z$ and $\overline{Z}$ applied
to obtain the correct continuum form.  Of course, such factors
are not needed to extract the correct mass from the large-time
limit of such Green's functions.

With this background, we can now discuss the work of the Tsukuba
group.  They emphasizes the importance of working with five
parameters, one more than the number determined in the Fermilab paper.
By introducing a fifth parameter, they are able to include an
additional field transformation which eliminates the parameter $\delta$
above, insuring a lattice action which will yield on-shell
quark propagators which take directly the continuum form.
This is useful for lattice perturbative calculations where such
on-shell quark propagators have meaning.

The Tsukuba group uses the field equations to derive their
results, not the approach using field transformations taken in
the Fermilab work and used in the present paper.  However,
there is no difference between these two methods because
one typically justifies the use of the equations of motion
when evaluating an on-shell amplitude by applying field
transformations in the path integral.  These two approaches
are formally equivalent in this situation.  The additional field
transformation of Eq.~\ref{eq:trans_2a} which permits us
to use $c_E=c_B$ can also be cast as a field equation implying
the same result.  Thus, we conclude that from both approaches
only three parameters are needed if an improved lattice action
is to yield continuum on-shell Green's functions, up to
the spinor transformations of Eqs.~\ref{eq:Z_factor_a} and
\ref{eq:Z_factor_b}.

Since our objective is to use the improved lattice action
to compute non-perturbative quantities, we do not benefit from
simplifying on-shell quark Green's functions.  However,
the field renormalization discussed above applies equally well
to composite spin-1/2 operators that might be used to create,
for example, a charmed baryon.  Since such a composite operator
will receive significant contributions from lattice-distorted
short-distances, the quantities $Z$ and $\delta$ appropriate for
such a physical heavy fermion will be different from a single
quark field and the Tsukuba choice of a fifth (now fourth)
parameter will not make $\delta$ vanish for the case of a
charmed baryon operator.  Fortunately, from a non-perturbative
perspective, the $Z$-factors above are relatively easy to
deal with.  They do not need to be known in advance and do
not effect the action used in a simulation.  Instead, they
can be easily determined {\it a posteriori} from the large
time behavior of the heavy baryon propagator and then used
elsewhere to accurately remove the lattice artifacts associated
with using that heavy quark composite field.

\section{Inductive transformation of the effective action}
\label{sec:induction}

In this section we study the complete continuum effective action
given in Eqs.~\ref{eq:eff}-\ref{eq:eff_1} whose coefficients are
polynomials of arbitrary order in $m_0a$, $(aD^0)^2$ and $g^2$.
We will use induction in the order of these polynomials to
demonstrate that by applying the field transformations of
Eqs.~\ref{eq:field_trans_0a}-\ref{eq:field_trans_2b} this general
effective action can be transformed to that given in
Eq.~\ref{eq:action_cont} where only the coefficients $m_r$,
$\zeta^c$ and $c^c_B=c^c_E$ are non-zero and functions of $m_0a$ and
$g^2$ alone.

The coefficient functions $B^{i,j}$ and $C^{i,j}$ appearing
in the original effective action of Eqs.~\ref{eq:eff}-\ref{eq:eff_1}
are polynomials of arbitrary order in the product $m_0a$, the
operator $(aD^0)^2$ and the gauge coupling $g^2$:
\begin{eqnarray}
B^{i,j}= \sum_{k,l,n} b_{k,l,n}^{i,j} (m_0a)^k(aD^0)^{2l} g^{2n} \nonumber \\
C^{i,j}= \sum_{k,l,n} c_{k,l,n}^{i,j} (m_0a)^k(aD^0)^{2l} g^{2n}.
\label{eq:coef_series2}
\end{eqnarray}
For later purposes it is important to recognize that only the usual terms
in the Dirac action will have non-zero coefficients in leading order:
\begin{equation}
b_{0,0,0}^{-1,1}=0, \quad b_{1,0,0}^{-1,1}=c_{0,0,0}^{-1,1}=2 b_{0,0,0}^{0,1}=1.
\label{eq:cont_tree}
\end{equation}

We will find it convenient to reorganize the sums in Eq.~\ref{eq:coef_series2}
collecting terms into homogenous polynomials of degree $N$ in the three variables,
$m_0a$, $(aD^0)^2$ and $g^2$:
\begin{eqnarray}
B^{i,j}= \sum_N b_N^{i,j}  \nonumber \\
C^{i,j}= \sum_N c_N^{i,j}. \label{eq:homo_series}
\label{eq:coef_series_1}
\end{eqnarray}
where $b_N^{i,j}$ and $c_N^{i,j}$ are such homogenous polynomials of degree $N$
in these three variables.  In terms of these polynomials, the character of the
tree-level, continuum limit of the standard Wilson action can be summarized
by the requirement that $b_0^{-1,1}=0$, $b_1^{-1,1}=m_0a$, and
$c_0^{-1,1}= 2 b_0^{0,1} = 1$ (equivalent to Eq.~\ref{eq:cont_tree}) and that
all the other $N=0$ coefficients $b_0^{i,j}$ and $c_0^{i,j}$ must vanish.

Equations~\ref{eq:eff}, \ref{eq:eff_-1}, \ref{eq:eff_0} and \ref{eq:eff_1}
are organized in increasing powers of the lattice spacing where we treat
$m$ and $D^0$ as order $1/a$ to accommodate the possibility that
$m \sim 1/a$.  However, it is important to bear in mind that the term
${\cal L}_{{\rm eff}, n}$ is characterized only by the lack of terms of
lower order in $a$ than $a^n$.  This term will necessarily contain
terms that are of higher order in $a$.   Commutators/anti-commutators have
been introduced into the definitions in Eqs.~\ref{eq:eff_-1}, \ref{eq:eff_0}
and \ref{eq:eff_1} in an attempt to organize these higher order terms.
The polynomials $B^{i,j}$ and $C^{i,j}$ above are labeled so the left
index indicates the order of the term in this scheme for RHQ power
counting, {\it e.g.} $O(a^{i})$ while the right index enumerates the
various terms that can occur in that order.  Note, we are using two
separate expansions.  One expansion is in powers of $a$, presuming that $m$
may be of order $1/a$.  The second is the expansion in the over-all order of
the three variables $m_0a$, $(aD^0)^2$ and $g^2$ were the term
$(m_0a)^k (aD^0)^{2l} g^{2n}$ is identified as of order $N=k+l+n$.

\subsection{Field Transformations}

As is discussed above, many of the unwanted terms in Eq.~\ref{eq:eff} have
no effect on physical states or fermion Green's functions evaluated on
the mass shell and can be removed by a redefinition of the fermion fields
$\psi$ and $\overline{\psi}$.  We will now make a series of such
transformations chosen to remove many of the terms that appear in the
Symanzik effective action of Eq.~\ref{eq:eff}.  The coefficients
of those terms that remain after these transformations are then presumed
to be potentially important lattice artifacts that should be eliminated
by an explicit choice of additional improvement terms in the underlying
lattice action.

The removal of these redundant terms is most easily analyzed in a series
of steps exploiting the ordering of the terms in Eq.~\ref{eq:eff}:
$O(1/a)$, $O(a^0)$, $O(a)$, {\it etc.} in the RHQ expansion.  We will first
make the large, $O(a^0)$, transformation of Eqs.~\ref{eq:field_trans_0a} and
\ref{eq:field_trans_0b} which we will be able to chose to return the $O(1/a)$
terms in ${\cal L}_{{\rm eff},-1}$ to the form found in the conventional
continuum action:
\begin{equation}
{\cal L}_{\rm sym} = \overline{\psi}'(\gamma^\mu D_\mu + m_r)\psi'.
\label{eq:eff_target}
\end{equation}
We will then consider the effect of both this $O(a^0)$ transformation as
well the most general $O(a)$ transformation given in Eqs.~\ref{eq:field_trans_1a}
and \ref{eq:field_trans_1b} on the $O(a^0)$ term ${\cal L}_{{\rm eff},0}$.
Finally, the effect of all three transformations, $O(a^0)$, $O(a)$ and the
$O(a^2)$ given in Eqs.~\ref{eq:field_trans_2a} and \ref{eq:field_trans_2b}
will be studied on the final term of interest, ${\cal L}_{{\rm eff},1}$.

\subsubsection{Redundant terms in ${\cal L}_{{\rm eff},-1}$}

In order to analyze the $O(1/a)$ terms in the Symanzik effective action,
we must consider the effects of a field transformation of $O(a^0)$ on
that action.  The most general such field transformation are given in
Eqs.~\ref{eq:field_trans_0a} and \ref{eq:field_trans_0b} and repeated here
for convenience, incorporating charge conjugation symmetry:
\begin{eqnarray}
\psi            &=& (1+R^{0,1} + a\gamma^0 D^0 S^{0,1})\psi'
\label{eq:field_trans_0a2} \\
\overline{\psi} &=& \overline{\psi}'(1+R^{0,1}
                     - a\gamma^0 \overleftarrow{D}^0 S^{0,1}).
\label{eq:field_trans_0b2}
\end{eqnarray}
The two functions $R^{0,1}$ and $S^{0,1}$ are polynomials of arbitrary
order in $m_0a$, $(aD^0)^2$ and $g^2$, similar to the coefficient functions
$B^{i,j}$ and $C^{i,j}$ of Eqs.~\ref{eq:eff_-1}-\ref{eq:eff_1} and
\ref{eq:coef_series2}.  These transformations are most easily analyzed if
we proceed in a systematic fashion, removing sequentially terms in
${\cal L}_{{\rm eff},-1}$ of increasing order $m_0a$, $(aD^0)^2$ and $g^2$
where, as above, we identify a term of the form $(m_0a)^k(aD^0)^{2l} g^{2n}$
as being of order $N=k+l+n$.  Reliance on such a formal expansion is a
standard approach to linearize the problem at hand, at the expense of
requiring that an inductive argument be created to deal with polynomials
of arbitrary order.

Specifically, we will achieve the general field transformation
described in Eqs.~\ref{eq:field_trans_0a2} and \ref{eq:field_trans_0b2}
by performing a sequence of simpler transformations where each
involves a homogenous polynomial of order $N$ in the three variables
$m_0a$, $(aD^0)^2$ and $g^2$:
\begin{eqnarray}
\psi            &=& (1+r_N^{0,1} + a\gamma^0 D^0 s_N^{0,1})\psi'
\label{eq:field_trans_0_Na} \\
\overline{\psi} &=& \overline{\psi}'(1+r_N^{0,1}
                     - a\gamma^0 \overleftarrow{D}^0 s_N^{0,1}).
\label{eq:field_trans_0_Nb}
\end{eqnarray}
Here the quantities $r_N^{i,j}$ and $s_N^{i,j}$ are homogenous polynomials
of order $N$ in the three variables $m_0a$, $(aD^0)^2$ and $g^2$.  The index
$i$ identifies the order of the term in the RHQ expansion and the index $j$
labels the specific operator appearing in the transformation.

\underline{Theorem} By proper choice of the transformation coefficients,
$r_N^{0,1}$ and $s_N^{0,1}$ it is possible to transform
${\cal L}_{\rm eff,-1}$ into the form:
\begin{equation}
{\cal L}_{\rm eff,-1} = \overline{\psi}'\{\gamma^0 D^0 + m_r\}\psi'
\label{eq:induct_0}
\end{equation}
where $m_r$ is a polynomial in the variables $m_0a$ and $g^2$.  This
theorem can be proven by induction in $N$.

\underline{Proof} To leading order in $N$, Eq.~\ref{eq:induct_0}
is satisfied without any transformation.  As observed above, the
coefficient of $\gamma^0 D^0$, $C^{-1,1} = c_0^{-1,1} =1$ to order
$N=0$.  Likewise at order $N=1$, the coefficient of $1/a$,
$B^{-1,1} = b_0^{-1,1} + b_1^{-1,1} = m_0a$ so that through order
$N=1$, $m_r=m$.   Thus, as the first step in our induction proof,
we note that $c_0^{-1,1} = 1$, $b_0^{-1,1}=0$ and
$b_1^{-1,1}=m_0a$.

Next we assume Eq.~\ref{eq:induct_0} is valid to order $N=N_0$
in the sense that after the previous $N_0$ steps, the resulting
coefficients in the Lagrangian ${\cal L}_{\rm eff,-1}$ of
Eq.~\ref{eq:eff_-1} obey: $C^{-1,1} = c_0^{-1,1} = 1$
and $B^{-1,1} = (m_ra)_{N_0+1}$.  Thus, we must attempt
to remove the next order terms in ${\cal L}_{\rm eff,-1}$:
\begin{equation}
{\cal L}_{\rm eff,-1}
    = \overline{\psi}\Bigl\{\gamma^0 D^0(1+c_{N_0+1}^{-1,1})
      + \frac{1}{a} \Bigl( (m_ra)_{N_0+1} + b_{N_0+2}^{-1,1} \Bigr)
                                                    \Bigr\}\psi.
\label{eq:induct_1}
\end{equation}
Here, by induction, $(m_ra)_{N_0+1}$ is assumed to be a polynomial
of order $N \le N_0+1$ in the variables $m_0a$ and $g^2$.  The
coefficients $b_{N_0+2}^{-1,1}$ and  $c_{N_0+1}^{-1,1}$ are
closely related to those appearing in Eqs.~\ref{eq:coef_series2}
and \ref{eq:coef_series_1}, differing only by the effects of the
field transformations made previously to achieve the form in
Eq.~\ref{eq:induct_1}:
\begin{eqnarray}
\psi      &=& \prod_{N=0}^{N_0}\Bigl(1+r_{N}^{0,1}
          + a\gamma^0 D^0 s_{N}^{0,1}\Bigr)\psi' \\
\overline{\psi} &=& \overline{\psi'}\prod_{N=0}^{N_0}\Bigl(1
          +r_{N}^{0,1} -a\gamma^0 \overleftarrow{D}^0 s_{N}^{0,1}\Bigr).
\label{eq:field_trans_gen}
\end{eqnarray}
Performing the next transformation of order $N_0+1$:
\begin{eqnarray}
\psi            &=& (1+r_{N_0+1}^{0,1}
                       + a\gamma^0 D^0 s_{N_0+1}^{0,1})\psi'
\label{eq:field_trans_0_1c} \\
\overline{\psi} &=& \overline{\psi'}(1+r_{N_0+1}^{0,1}
       - a\gamma^0 \overleftarrow{D}^0 s_{N_0+1}^{0,1}).
\label{eq:field_trans_0_1d}
\end{eqnarray}
${\cal L}_{\rm eff,-1}$ of Eq.~\ref{eq:induct_1} becomes:
\begin{eqnarray}
{\cal L}_{\rm eff,-1}
   &=& \overline{\psi'}\Bigl\{\gamma^0 D^0\bigl[1+c_{N_0+1}^{-1,1}
       + 2r_{N_0+1}^{0,1} \bigr] \nonumber \\
     &&+   \frac{1}{a} \bigl[ (m_ra)_{N_0+1} + 2 b_{N_0+2}^{-1,1}
            + 2 (aD^0)^2 s_{N_0+1}^{0,1} \bigr] \Bigr\}\psi'.
\label{eq:induct_2}
\end{eqnarray}
Thus, we can establish our theorem to order $N_0+1$ if we
require:
\begin{equation}
c_{N_0+1}^{-1,1} + 2 r_{N_0+1}^{0,1} = 0
\label{eq:induct_3}
\end{equation}
and choose $s_{N_0+1}^{0,1}$ to remove the $(aD^0)^{2N}$ terms for
$1 \le N \le N_0+2$ from the coefficient $b_{N_0+2}^{-1,1}$ so that
the definition
\begin{eqnarray}
(m_ra)_{N_0+2} &=& (m_ra)_{N_0+1} + b_{N_0+2}^{-1,1}
             + 2(aD^0)^2 s_{N_0+1}^{0,1}
\label{eq:induct_4}
\end{eqnarray}
will contain no $(aD^0)^2$ terms as required.

Following this inductive procedure, we are thus able to express the
$O(1/a)$ Symanzik Lagrangian in the standard continuum form.  Only
the mass parameter $m_r$ must be tuned by an appropriate choice of
lattice action to agree with the mass of the heavy quark which this
Lagrangian is intended to describe.

\subsubsection{Redundant terms in ${\cal L}_{{\rm eff},0}$}

The order $a^0$, Symanzik effective Lagrangian, ${\cal L}_{{\rm eff},0}$
is altered by two sorts of field transformations.  The first is the
$O(a^0)$ transformations discussed above.  The second are the $O(a)$
transformations of Eqs.~\ref{eq:field_trans_1a} and \ref{eq:field_trans_1b}
which act on ${\cal L}_{{\rm eff},-1}$ and generate terms of the type
which appear in ${\cal L}_{{\rm eff},0}$.  Including the constraints of
charge conjugation symmetry these $O(a)$ transformations can be written:
\begin{eqnarray}
\psi            &=& \Bigl(1+ a \vec\gamma \vec D R^{1,1}
                 + a^2[\vec\gamma \vec D,\gamma^0 D^0] S^{1,1}\Bigr)\psi'
\label{eq:field_trans_1_1a} \\
\overline{\psi} &=& \overline{\psi}'\Bigl((1
                - a \vec\gamma \overleftarrow{D} R^{1,1}
                - a^2[\vec\gamma \overleftarrow{D},\gamma^0
                           \overleftarrow{D}^0] S^{1,1}\Bigr).
\label{eq:field_trans_1_1b}
\end{eqnarray}
As in the previous discussion, it will be convenient to view the
coefficient functions $R^{1,1}$ and $S^{1,1}$ as a sum of homogenous
polynomials in the three variables $m_0a$, $(aD^0)^2$ and $g^2$:
\begin{eqnarray}
R^{i,j} &=& \sum_N r_N^{i,j} \quad\quad S^{i,j} = \sum_N s_N^{i,j}
\label{eq:poly_exp_1}
\end{eqnarray}
Again, we will proceed inductively to prove the following result:

\underline{Theorem} By proper choice of the transformation
coefficients, $r_N^{1,1}$ and $s_N^{1,1}$, it is possible to
transform ${\cal L}_{\rm eff}$ so that ${\cal L}_{\rm eff,0}$ takes
the form:
\begin{equation}
{\cal L}_{\rm eff,0}
            = \overline{\psi}\vec\gamma \vec D \psi.
\label{eq:induct_0b}
\end{equation}

\underline{Proof} This result is automatically valid to order $N=0$
which is the case of the tree-level Lagrangian with $b_0^{0,1}=1/2$
and $c_0^{0,1}=0$.  Next, assume the inductive hypothesis that when
working to order $N_0$ we are able to simplify ${\cal L}_{{\rm eff},0}$
so that all terms of order $N_0+1$ and lower take the form:
\begin{equation}
{\cal L}_{{\rm eff},0} = \overline{\psi} \Bigl\{\vec\gamma \vec D, (1/2
         + b_{N_0+1}^{0,1})\Bigr\} \psi.
\label{eq:eff_0_N_0a}
\end{equation}
(Recall that the $C^{0,1}$ term in Eq.~\ref{eq:eff_0} vanishes when
charge conjugation symmetry is imposed.)

We will now apply the transformations of order $N_0+1$ given in
Eqs.~\ref{eq:field_trans_0_1c} and \ref{eq:field_trans_0_1d} and
those in Eqs.~\ref{eq:field_trans_1_1a} and \ref{eq:field_trans_1_1b}
specialized to the polynomials of order $N_0$,
\begin{eqnarray}
\psi         &=& \Bigl(1+ a \vec\gamma \vec D r_{N_0}^{1,1}
              + a^2[\vec\gamma \vec D,\gamma^0 D^0] s_{N_0}^{1,1}\Bigr)\psi'
\label{eq:field_trans_1_Na} \\
\overline{\psi}
             &=& \overline{\psi'}\Bigl(1
               -a r_{N_0}^{1,1}\, \vec\gamma \overleftarrow D
          - a^2 s_{N_0}^{1,1} [\vec\gamma \overleftarrow{D},\gamma^0 \overleftarrow{D}^0]
                  \Bigr),
\label{eq:field_trans_1_Nb}
\end{eqnarray}
to ${\cal L}_{{\rm eff},-1}+{\cal L}_{{\rm eff},0}$.

These transformations yield ${\cal L}_{{\rm eff},0}$ of the following form:
\begin{eqnarray}
{\cal L}_{{\rm eff},0} &=& \overline{\psi'} \Bigl\{\vec\gamma \vec D, \Bigl(\frac{1}{2}
                         + b_{N_0+1}^{0,1} + r^{0,1}_{N_0+1} + m_0 a\,r^{1,1}_{s,N_0}
                      -(aD^0)^2s^{1,1}_{N_0} \Bigr)\Bigr\}\psi'.
\label{eq:eff_0_N_0}
\end{eqnarray}
Since the difference between the coefficient of $\gamma^0 D^0$ which
has now been set to one and that of $\vec \gamma \vec D$ must vanish
when the anisotropic effects of the special treatment of $m_0a$ and $D^0$
are absent, the combination $b_{N_0+1}^{0,1} + r^{0,1}_{N_0+1}$ must be
proportional to a linear combination of $m_0a$ and $(aD^0)^2$ and can
therefore be completely canceled by an appropriate choice of the terms
$m_0 a\, r^{1,1}_{N_0}$ and $-2(aD^0)^2s^{1,1}_{N_0}$, completing our
inductive proof.

As the preceding discussion reveals, our inductive approach to
determining the redundant parameters in ${\cal L}_{\rm eff}$
requires that both the order $a^0$ and order $a^1$ field transformations
are to be applied at the same time so that a common inductive step
is taken to show that the desired form will hold at order $N_0+1$
provided it holds at order $N_0$.  It is in this sense that we are
combining the order $a^0$ and $a^1$ transformations in
Eq.~\ref{eq:eff_0_N_0}.

\subsubsection{Redundant terms in ${\cal L}_{{\rm eff},1}$}

The last step in this discussion is an analysis of the freedom to
simplify the terms of order $a$ in ${\cal L}_{\rm eff}$, {\it i.e.}
${\cal L}_{\rm eff,1}$.  These can be effected by three different
sorts of field transformations:  transformations of order $a^0$ acting on
${\cal L}_{\rm eff,1}$, transformations of order $a$ acting on
${\cal L}_{\rm eff,0}$ and transformations of order $a^2$ acting on
${\cal L}_{\rm eff,-1}$.  We will again state our result in the
form of a theorem to be proven by induction in the order of the
polynomials appearing in ${\cal L}_{\rm eff,1}$:

\underline{Theorem}  By an appropriate field transformation
${\cal L}_{\rm eff,1}$ can be cast in the form:
\begin{equation}
{\cal L}_{\rm eff,1} = -\overline{\psi} c_{P}\Bigl\{
    \frac{1}{8} [\gamma^i,\gamma^j][D^i,D^j] + \frac{1}{4}[\gamma^i,\gamma^0][D^i,D^0]\Bigr\}
                            \psi
\label{eq:induct_0c}
\end{equation}
where $c_{P}=B^{1,2}=B^{1,3}/2$ is a polynomial in $m_0a$ and $g^2$ only.

\underline{Proof}  We begin by observing that to order $N=0$
Eq.~\ref{eq:induct_0c} is automatically obeyed with $c_{P}=-8 B_{N=0}^{1,2}=1$,
the original, tree-level result of Sheikholeslami and Wohlert.  Next we
assume that this is true to order $N_0$ so that to order $N_0+1$,
${\cal L}_{\rm eff,1}$ takes the form:
\begin{eqnarray}
{\cal L}_{\rm eff,1}
  &=&  a \overline{\psi}\Biggl\{\vec D^2 b_{N_0+1}^{1,1}
     + a\{\vec D^2, \gamma^0 D^0\} c_{N_0+1}^{1,1}
                                            \label{eq:induct_1c} \\
  &&\quad+ [\gamma^i, \gamma^j] [D^i,D^j]\Bigl(-\frac{1}{8}(c_{P})_{N_0}+ b_{N_0+1}^{1,2}\Bigr)
                                                                 \nonumber \\
  &&\quad+ a\{[\gamma^i, \gamma^j][D^i,D^j], \gamma^0 D^0 \}c_{N_0+1}^{1,2}
                                                                 \nonumber \\
  &&\quad+ [\gamma^i, \gamma^0] [D^i,D^0]\Bigl(-\frac{1}{4}(c_{P})_{N_0}+ b_{N_0+1}^{1,3}\Bigr)
                                                        \Biggr\}\psi. \nonumber
\end{eqnarray}

We will now attempt to remove the redundant terms in Eq.~\ref{eq:induct_1c}
by the following three field transformations.  The first is the $O(a^0)$
transformations of Eqs.~\ref{eq:field_trans_0_Na} and \ref{eq:field_trans_0_Nb}
that involve polynomials in $m_0a$, $(aD^0)^2$ and $g^2$ of combined order
$N=N_0+1$.  These $O(a^0)$ transformations will have the following $O(a)$ effects.
When acting on ${\cal L}_{\rm eff,-1}$, these $O(a^0)$ transformations produce
only terms of $O(1/a)$.  No terms of $O(a^0)$ or $O(a)$ are produced.  If
these $O(a^0)$ transformations act on ${\cal L}_{\rm eff,0}$ both terms of
$O(a^0)$ and of $O(a)$ are created.  Those of $O(a^0)$ appear in
Eq.~\ref{eq:eff_0_N_0} and have been removed by the transformations of order
$a^1$.  The terms of $O(a)$ which will effect ${\cal L}_{\rm eff,1}$ take the
following form:
\begin{eqnarray}
\Delta{\cal L}_{\rm eff,1}^{0,0} &=& a\overline{\psi}
             \{\vec \gamma \vec D, \gamma^0 D^0\}\Bigl(s_{s,N_0+1}^{0,1}
              +2(aD^0)^2 \frac{s_{s,N_0+1}^{0,1}}{\partial((aD^0)^2)}\Bigr)\psi
\label{eq:delta_0_0}
\end{eqnarray}
Here the $i,j$ superscript on $\Delta{\cal L}_{\rm eff,1}^{i,j}$
identifies this expression as the change in ${\cal L}_{\rm eff,1}$
coming from applying a transformation of order $a^i$ to ${\cal L}_{\rm eff,j}$.

Next we should consider the effect of this $O(a^0)$ transformation on the
$O(a)$ Lagrangian ${\cal L}_{\rm eff,1}$.  However, since we will not
need to use the effects of this transformation on ${\cal L}_{\rm eff,1}$,
we will assume that its effects have already be taken into account in
the coefficients $b_{N_0+1}^{1,j}$ and $c_{N_0+1}^{1,j}$ that appear in
Eq.~\ref{eq:induct_1c}.

Having completely accounted for the effects on ${\cal L}_{\rm eff}$
of the transformations of $O(a^0)$ given in Eqs.~\ref{eq:field_trans_0_1c}
and \ref{eq:field_trans_0_1d}, we will now consider the transformations of
$O(a)$ given in Eqs.~\ref{eq:field_trans_1_Na} and \ref{eq:field_trans_1_Nb}.
First as they act on ${\cal L}_{\rm eff, -1}$ they will produce the
following changes in ${\cal L}_{\rm eff,1}$:
\begin{eqnarray}
\Delta{\cal L}_{\rm eff,1}^{1,-1} &=& \frac{a}{2}\overline{\psi}
       [\gamma^i,\gamma^0][ D^i,D^0] r_{N_0+1}^{1,1} \psi
\label{eq:delta_1_-1}
\end{eqnarray}
Note, this term was generated from the $\gamma^0 D^0$ term
in ${\cal L}_{\rm eff, -1}$.  No terms of order $a$ are produced from
the $m$ term.

The next case to consider is the effect of these transformations of
$O(a)$ on ${\cal L}_{\rm eff, 0}$.  The resulting changes to
${\cal L}_{\rm eff,1}$ are:
\begin{eqnarray}
\Delta{\cal L}_{\rm eff,1}^{1,0} &=& \overline{\psi}\Biggl\{
     a\Bigl(2 \vec D^2 + \frac{1}{2}[\gamma^i,\gamma^j][D^i,D^j]\Bigr)
                    r^{1,1}_{N_0+1}
\label{eq:delta_1_0} \\
    && + a\Bigl\{\Bigl(2 \vec D^2 + \frac{1}{2}[\gamma^i,\gamma^j][D^i,D^j]\Bigr),
           a\gamma^0 D^0\Bigr\}s^{1,1}_{N_0+1}
     \Biggr\}\psi.  \nonumber
\end{eqnarray}

The final $O(a)$ effects to consider are those of transformations of
$O(a^2)$ acting on ${\cal L}_{\rm eff, -1}$.  If Eqs.~\ref{eq:field_trans_2a}
and \ref{eq:field_trans_2b} are specialized to respect charge conjugation
symmetry, the relevant $O(a^2)$ field transformations can be written:
\begin{eqnarray}
\psi            &=& \Biggl(1+ a^2 \vec D^2 r_{N_0+1}^{2,1}
                   + a^2\{\vec D^2,a\gamma^0 D^0\}s_{N_0}^{2,1}
\label{eq:field_trans_2_Na} \\
                 &+&a^2[\gamma^i, \gamma^j][D^i, D^j] r_{N_0+1}^{2,2}
\nonumber \\
    &+&a^2\Bigl\{[\gamma^i, \gamma^j][D^i, D^j],a\gamma^0 D^0\Bigr\}s_{N_0}^{2,2}
\nonumber \\
   &+&a^2[\gamma^i, \gamma^0][D^i, D^0] r_{N_0}^{2,3}
\nonumber \\
   &+&a^2\Bigl[[\gamma^i, \gamma^0][D^i, D^0],a\gamma^0 D^0\Bigr]s_{N_0}^{2,3}
\Biggr)\psi' \nonumber \\
\overline{\psi} &=& \overline{\psi}'\Biggl(1
                   + a^2 \overleftarrow{D}^2 r_{N_0+1}^{2,1}
 - a^2\{\overleftarrow{D}^2,a\gamma^0 \overleftarrow{D}^0\} s_{N_0}^{2,1}
\label{eq:field_trans_2_Nb} \\
                &+&a^2[\gamma^i, \gamma^j][\overleftarrow{D}^i, \overleftarrow{D}^j]
                            r_{N_0+1}^{2,2}
\nonumber \\
        &-&a^2\Bigl\{[\gamma^i, \gamma^j][\overleftarrow{D}^i, \overleftarrow{D}^j],
               a\gamma^0 \overleftarrow{D}^0\Bigr\} s_{N_0}^{2,2}
\nonumber \\
        &+&a^2[\gamma^i, \gamma^0][\overleftarrow{D}^i, \overleftarrow{D}^0]
                 r_{N_0}^{2,3}
\nonumber \\
   &+& a^2\Bigl[ [\gamma^i, \gamma^0][\overleftarrow{D}^i, \overleftarrow{D}^0],
          a\gamma^0 \overleftarrow{D}^0\Bigr] s_{N_0}^{2,3} \Biggr).
\nonumber
\end{eqnarray}
The resulting $O(a)$ terms are:
\begin{eqnarray}
\Delta{\cal L}_{\rm eff,1}^{2,-1} &=& \overline{\psi}\Biggl\{
     a \vec D^2\Bigl( 2m_0a\,r_{N_0+1}^{2,1} +4(aD^0)^2 s_{N_0}^{2,1}\Bigr)
 \label{eq:delta_2_-1} \\
      && +a\{\vec D^2,a\gamma^0 D^0\}( 2m_0a\, s_{N_0}^{2,1} +r_{N_0+1}^{2,1})
\nonumber \\
      && +a[\gamma^i,\gamma^j][D^i,D^j]\Bigl( 2m_0a\,r_{N_0+1}^{2,2}
                           +4(aD^0)^2s_{N_0}^{2,2}\Bigr)
\nonumber \\
      && +a\{[\gamma^i,\gamma^j][D^i,D^j],\gamma^0D^0\}\Bigl(
                   2m_0a\, s_{N_0}^{2,2} +r_{N_0+1}^{2,2}\Bigr)
\nonumber \\
      && +a[\gamma^i,\gamma^0][D^i,D^0]\Bigl( 2m_0a\, r_{N_0+1}^{2,3}
                         -4(aD^0)^2s_{N_0}^{2,3}\Bigr) \Biggr\}\psi.
\nonumber
\end{eqnarray}

We can now combine the $O(a)$ terms created by these three field transformations
with those already present in Eq.~\ref{eq:induct_1c}.  We will do this by
considering in turn each of the three types of operators appear in
Eq.~\ref{eq:induct_1c} with coefficients whose right hand superscript is
$1 \le j \le 3$, which we will denote ${\cal L}_{\rm eff,1}^{(j=1,2,3)}$.

We first examine ${\cal L}_{\rm eff,1}^{(1)}$ constructed by collecting terms
from Eqs.~\ref{eq:induct_1c}, \ref{eq:delta_1_0} and \ref{eq:delta_2_-1}:
\begin{eqnarray}
{\cal L}_{\rm eff,1}^{(1)}
    &=& a \overline{\psi}\Biggl\{\vec D^2 \Bigl(b_{N_0+1}^{1,1}
                 +2r^{1,1}_{N_0+1} + 2m_0a\,r_{N_0+1}^{2,1}
                 +4(aD^0)^2s_{N_0}^{2,1} \Bigr)
\label{eq:collected_1_1} \\
     &&+ a\{\vec D^2, \gamma^0 D^0\} \Bigl( c_{N_0+1}^{1,1}
                 +2s^{1,1}_{N_0+1}+ 2m_0a\, s_{N_0}^{2,1}
                 +r_{N_0+1}^{2,1}\Bigr) \Biggr\}\psi. \nonumber
\end{eqnarray}
Since we will adopt the usual conventions of fixing the spatial Wilson term
in the lattice action to have normalization 1, we will adjust $r^{1,1}_{N_0+1}$
appearing above to remove the corresponding $\vec D^2$ term.  This implies
that we cannot make the choice described in the previous discussion of
${\cal L}_{\rm eff,0}$ to set the coefficient of $\vec \gamma \vec D$ to
one, which was also accomplished by a different choice of these same
coefficients, see Eq.~\ref{eq:eff_0_N_0} and following.  The second term
of the form $a\{\vec D^2,a \gamma^0 D^0\}$ in Eq.~\ref{eq:collected_1_1}
will be removed the choice of $r_{N_0+1}^{2,1}$.

We next examine ${\cal L}_{\rm eff,1}^{(2)}$ constructed by collecting terms
from Eqs.~\ref{eq:induct_1c}, \ref{eq:delta_1_0} and \ref{eq:delta_2_-1}:
\begin{eqnarray}
{\cal L}_{\rm eff,1}^{(2)}
    &=& a \overline{\psi}\Biggl\{ a[\gamma^i,\gamma^j][D^i,D^j]
      \Bigl(-\frac{1}{8}(c^c_{P})_{N_0} + b_{N_0+1}^{1,2} + r^{1,1}_{N_0+1}
+ 2m_0a\,r_{N_0+1}^{2,2} +4(aD^0)^2 s_{N_0}^{2,2} \Bigr) \nonumber\\
     &&+ a \Bigl\{[\gamma^i,\gamma^j][D^i,D^j],a\gamma^0 D^0\Bigr\}\Bigl\{
        c_{N_0+1}^{1,2} + \frac{1}{2} s^{1,1}_{N_0+1}
+ 2m_0a\, s_{N_0}^{2,2} +r_{N_0+1}^{2,2} \Bigr\}
                                      \Biggr\}\psi. \label{eq:collected_1_2}
\end{eqnarray}
Since the coefficient $r^{1,1}_{N_0+1}$ has already been used to remove
the $\vec D^2$ term and the coefficient $r_{N_0+1}^{2,2}$ will be used
below, we have only the freedom to adjust the combination $s_{N_0}^{2,2}$
to remove the terms proportional to $(aD^0)^2$ for the coefficient of
$a[\gamma^i,\gamma^j][D^i,D^j]$.  Thus, the parameter $c_{P}$ will
require mass-dependent tuning.  However, the second term,
$a \{[\gamma^i,\gamma^j][D^i,D^j],a\gamma^0 D^0\}$ can be entirely
removed by a choice of $r_{N_0+1}^{2,2}$.

Finally we consider the term ${\cal L}_{\rm eff,1}^{(3)}$ constructed by
collecting terms from Eqs.~\ref{eq:induct_1c}, \ref{eq:delta_0_0},
\ref{eq:delta_1_-1} and \ref{eq:delta_2_-1}:
\begin{eqnarray}
{\cal L}_{\rm eff,1}^{(3)}
    &=& a \overline{\psi} a[\gamma^i,\gamma^0][D^i,D^0]\Bigl\{
          -\frac{1}{4}(c_{P})_{N_0}+b_{N_0+1}^{1,3}+s_{s,N_0+1}^{0,1}
\label{eq:collected_1_3} \\
    &&  +2(aD^0)^2 \frac{s_{s,N_0+1}^{0,1}}{\partial((aD^0)^2)}
    +\frac{1}{2}r_{N_0+1} +2 m_0a\, r_{s,N_0+1}^{2,3} -4(aD^0)^2s_{N_0}^{2,3}
                                      \Bigr\}\psi. \nonumber
\end{eqnarray}
We can now exploit the freedom to choose the coefficient $s_{N_0}^{2,3}$
to remove the terms containing $(aD^0)^{2l}$ from the coefficient of
$[\gamma^i,\gamma^0][D^i,D^0]$ and can determine $m_0a\, r_{N_0}^{2,3}$
to set this coefficient equal to that of $[\gamma^i,\gamma^j][D^i,D^j]/2$
since their difference must be proportional to $m_0a$.

Thus, we have shown that with the proper choice of field transformations
to remove redundant terms, the general Symanzik action, invariant under
axis reversal and charge conjugation will contain only three independent
parameters.  This result is summarized in Table~\ref{tab:improvement} where
the various field transformations and the terms which they eliminate are
listed.  We conclude that a lattice calculation accurate through order
$|\vec p a|$ and to arbitrary order in $ma$ requires the determination of
the three parameters $m_0$, $\zeta$ and $c_P$ appearing in the improved
lattice action.

\section{Tree-level results}
\label{sec:tree}
%section giving tree-level discussion of propagator and vertex

In order to investigate the number of required parameters further, we have
carried out a tree-level calculation of both the quark propagator and the quark
gluon vertex for a general, heavy-quark lattice Lagrangian, but evaluated
in the limit $|\vec p a| \ll 1$.   We begin with the general lattice action
given in Eq.~\ref{eq:action_lat} which depends on six parameters, $m_0$, $\zeta$,
$r_s$, $r_t$, $c_B$ and $c_E$.  We then demonstrate that a continuum
result can be obtained on-shell, accurate through $O(|\vec p a|)$ and
to all orders in $m_r a$ by adjusting only the expected three parameters
$m_0$, $c_P \equiv c_B = c_E$ and $\zeta$ while at the same time performing
a simple $4 \times 4$ matrix rotation on the Dirac spinors.

The presence of hyperbolic trigonometric functions in the Minkowski-space
lattice propagator makes the algebra in this section somewhat complex.  This
complexity is compounded by the approximation $|\vec p a| \ll 1$ which is
being made to functions of the two variables $|\vec p a|$ and $m_r a$.
Depending on the size of the second variable $m_r a$, the treatment of the
quantity $|\vec p a|$ can be quite different.  It is natural to divide the
possible values of $m_r a$ into two regions.  In the first region
$m_r a \ll 1$, and we have the kinematics of standard, light fermions.   In
this case we cannot neglect $|\vec p|/m_r$ but can treat $m_r a$ as a small
parameter.  In second region we assume $p \ll m_r$.  Here we cannot
neglect errors of order $m_ra$ but can treat $|\vec p|/m_r$ as small.
Since these two regions have a non-vanishing overlap,
$|\vec p| \ll m_r \ll 1/a$, we will be able to demonstrate that the tree
level amplitudes are consistent with our treatment for all values of
$m_r a$ if we are able to provide satisfactory bounds on the errors
in both of these two regions.

We propose to do this as follows.  First we introduce a small parameter
$\epsilon = |\vec p a|$.  Our objective is to show that at tree level,
working with the improved, 3-parameter action and an appropriate
$4 \times 4$ spinor transformation matrix, we can reproduce continuum
results up to errors of order $\epsilon^2$.  We will divide the range
of values of $m_r a$ into two non-overlapping regions.  In the first,
Region I, we require $m_r a \le \sqrt{\epsilon}$.  Here we can Taylor expand
in the parameter $m_r a$ but must control errors up to order $(m_r a)^4$.
Region II corresponds to the remaining range of $m_r$:
$\sqrt{\epsilon} < m_r a$.  Now we can expand in
$|\vec p|/m_r = |\vec p| a/(m_ra)\le \sqrt{\epsilon}$ but must therefore
work up to $O((\vec p/m_r)^4)$.

\subsection{Momentum-dependent energy}

The quark wave function renormalization constant $Z_q$ and the
parameters $m_0$, $\zeta$ and $r_s$ can be constrained by demanding that
the lattice quark propagator $G_q(p)$ derived from Eq.(1) should reproduce
the relativistic form
\begin{equation}
G_q(p_0,p_i)=\frac{1}{Z_q}\frac{-i\gamma^0p_0-i\vec\gamma\cdot\vec p+m_r}
                                                  {p_0^2+\vec p^2+m_r^2}
    +(\mbox{non-pole terms}) +O\Bigl((p_ia)^2\Bigr)
\label{eq:cont_prop}
\end{equation}
at the heavy quark pole in the limit $|p_i a| \ll 1$.  The location of the
pole in the tree-level lattice propagator is that value of $p_0$ at which
the inverse propagator vanishes:
\begin{eqnarray}
aG^{-1}_q(p_0,p_i)=i\gamma^0\sin(p_0a)+i\zeta\sum_i\gamma^i\sin(p_ia)+m_0a \nonumber \\
+r_t\Bigl(1-\cos(p_0a)\Bigr)+r_s\sum_i\Bigl(1-\cos(p_ia)\Bigr)=0.
\label{eq:pole}
\end{eqnarray}
By first examining the simplest case of zero spatial momenta, $p_i=0$, we
can obtain equations for $m_0$ and $Z_q$:
\begin{equation}
m_ra=\ln\Biggl(\frac{m_0a+r_t+\sqrt{(m_0a)^2+2r_tm_0a+1}}{1+r_t}\Biggr)
\label{eq:pole_zp}
\end{equation}
\begin{equation}
Z_q=\cosh(m_ra)+r_t\sinh(m_ra)
\label{eq:Z_q}
\end{equation}
To obtain constraints on $\zeta$ and $r_s$ we need to examine the case
of finite spatial momentum.

From the dispersion relation $p_0=i\sqrt{m_r^2+\vec p^2}$ which
we would like to reproduce, we can get a relationship between
$r_s$ and $\zeta$. Starting from Eq.~\ref{eq:pole} and defining a new
variable $\tilde{p}_0 \equiv -ip_0$, we obtain:
\begin{equation}
(r_t^2-1)\cosh^2(\tilde{p}_0a)-2r_tB\cosh(\tilde{p}_0a)+1+\zeta^2\sin^2(p_ia)+B^2=0
\end{equation}
where $B=r_t+m_0a+r_s\sum_i(1-\cos(p_ia))$. Neglecting quantities of order
$O((p_ia)^2)$ and higher, the two roots of the quadratic equation for
$\cosh(\tilde{p}_0a)$ can be written:
\begin{equation}
R_{\pm}=\frac{r_tB\pm\sqrt{r_t^2B^2-(r_t^2-1)(1+B^2+\zeta^2\vec p^2)}}{r_t^2-1}
\end{equation}
Here we choose $R_{-}$ as the physical root since $R_{+}$ goes to infinity
when $r_t \rightarrow 1$.  After we substitute the expression for $B$ into
the $R_{-}$ and expand to first order in the quantity $(\vec p a)^2$, we find:

\begin{eqnarray}
\cosh{(\tilde{p}_0a)}&=&\frac{r_t(m_0a+r_t)}{r_t^2-1}+\frac{r_tr_s(p_ia)^2/2}{r_t^2-1}
             \\
&-&\frac{\sqrt{[(m_0a+r_t)^2-(r_t^2-1)]+[r_s(m_0+r_t)-\zeta^2(r_t^2-1)](p_ia)^2}}{r_t^2-1}
   \nonumber \\
&=&\frac{r_t(m_0a+r_t)-\sqrt{(m_0a+r_t)^2-(r_t^2-1)}}{r_t^2-1}
   \nonumber \\
&+&\{\frac{r_tr_s}{2(r_t^2-1)}-\frac{r_s(m_0a+r_t)
  -\zeta^2(r_t^2-1)}{2(r_t^2-1)\sqrt{(m_0a+r_t)^2-(r_t^2-1)}}\}(p_ia)^2
             \\
&=&\cosh(m_ra)+\frac{r_s\sinh(m_ra)+\zeta^2}{2(r_t\sinh(m_ra)+\cosh(m_ra))}(p_ia)^2
\label{eq:pole_nzp}
\end{eqnarray}
where the last line is obtained using Eq.~\ref{eq:pole_zp}.

Equation~\ref{eq:pole_nzp} can be rewritten in the suggestive form:
\begin{equation}
\tilde{p}_0a = \sinh^{-1}\Biggl\{\sqrt{ \sinh^2(m_r a)
   + (\vec p a)^2 \cosh(m_r a)\frac{r_s\sinh(m_ra)+\zeta^2}{r_t\sinh(m_ra)+\cosh(m_ra)}}\Biggr\}.
\label{eq:pole_nzp2}
\end{equation}
If $m_r a \ll 1$ then $\sinh(z)$ and $\sinh^{-1}(z)$ can both be replaced by
$z$ and Eq.~\ref{eq:pole_nzp2} gives the usual relativistic dispersion relation
if we set $\zeta =1$.  If $m_r a$ is sufficiently large that this approximation
to $\sinh(z)$ and $\sinh^{-1}(z)$ is a poor one, then we can expand the square
root in Eq.~\ref{eq:pole_nzp2} to first order in $(\vec p a)^2$ and obtain the
result:
\begin{equation}
\tilde{p}_0a = m_r a
  + \frac{(\vec p a)^2}{2\sinh{m_r a}}\,\frac{r_s\sinh(m_ra)+\zeta^2}
                                      {r_t\sinh(m_ra)+\cosh(m_ra)}.
\label{eq:pole_nzp3}
\end{equation}
Thus, we will obtain the correct dispersion relation in both cases
if we require:
\begin{equation}
r_s\sinh(m_ra)+\zeta^2
            =\frac{\sinh(m_ra)}{m_ra}\Bigl(r_t\sinh(m_ra)+\cosh(m_ra)\Bigr).
\label{eq:r_s-zeta_relation}
\end{equation}

As discussed above, we can establish the equivalence of Eq.~\ref{eq:pole_nzp2}
to the usual dispersion relation
\begin{equation}
\tilde{p}_0a =
\sinh^{-1} \Biggl \{ \sqrt{\sinh^2(m_r a)
            +\frac{\sinh(m_r a)\cosh(m_r a)}{m_r a} (\vec p a)^2 }\Biggr\}
=\sqrt{(m_r a)^2 + (\vec p a)^2}
\label{eq:pole_nzp4}
\end{equation}
up to relative errors of order $(\vec p a)^2 \equiv \epsilon^2$ for all
values of $m_r a$ by showing it to holds to this accuracy in the two regions
$m_r a \le \sqrt{\epsilon}$ (region I) and $\sqrt{\epsilon} < m_r a$
(region II).  Here the left-hand equality in Eq.~\ref{eq:pole_nzp4} is
simply Eq.~\ref{eq:pole_nzp2} with the constraint in
Eq.~\ref{eq:r_s-zeta_relation} imposed.  Establishing that the right-hand
equality holds without errors larger than $O(\epsilon^2)$ requires in
Region I that we examine to next-leading order a Taylor series expansion
in the variables $(m_r a)^4/((m_r a)^2+(\vec p a)^2)$ and $(m_r a)^2$.
In Region II, we need only continue the expansion in $(\vec pa/m_r a)^2$
begun in Eq.~\ref{eq:pole_nzp3} to demonstrate that this second equality
holds up to relative errors of order $(\vec p a/m_r a)^4 \sim \epsilon^2$.
While straight-forward, some care is required to verify that the $m_r a$-dependent
coefficient of $(\vec p a/m_r a)^4$ is bounded throughout the region
$\epsilon \le m_r a$.

Thus, we can reproduce the correct momentum dependence of the
heavy quark energy if and only if the parameters $r_s$ and $\zeta$
satisfy the relationship in Eq.~\ref{eq:r_s-zeta_relation}.

\subsection{Propagator spinor structure}

Without a second constraint which then determines both $r_s$ and
$\zeta$, the general action under consideration will not reproduce the
correct spinor structure for the propagator.  However, as discussed
earlier, we can also achieve the conventional, on-shell spinor structure
for the propagator by applying a simple matrix transformation to the
on-shell spinor fields  even if we have chosen an arbitrary $r_s$ and
an appropriate value for $\zeta(r_s)$ so that Eq.~\ref{eq:r_s-zeta_relation}
is obeyed.  If we adopt this approach then we have freedom to choose
$r_s$ in the action for convenience, {\it e.g.} $r_s=\zeta$, thereby
reducing the number of parameters in the action by one.

We begin by examining the matrix form of the propagator as presently
determined:
\begin{equation}
aG_q(p_0,p_i)=\frac{-i\gamma^0\sin(p_0a)
-i\zeta\vec\gamma\cdot\vec p\, a + F}{\sin^2(p_0 a)+\zeta^2(\vec p a)^2+F^2}
\label{eq:matrix_res}
\end{equation}
where $F$ is given by
\begin{equation}
F = m_0 a+r_t(1-\cos(p_0a))+\frac{r_s}{2}(\vec p a)^2.
\end{equation}
We will now try to find a pair of $4 \times 4$ spinor matrices $U_L(\vec p)$
and $U_R(\vec p)$ able to transform the matrix in the numerator of
Eq.~\ref{eq:matrix_res} to the correct one:
\begin{eqnarray}
U_L(\vec p)\frac{-i\gamma^0\sin(p_0a)-i\zeta\gamma^i\sin(p_ia)+F}
{\sin^2(p_0a)+\zeta^2\sum_i\sin^2(p_ia)+F^2}\ U_R(\vec p)\nonumber\\
\approx \frac{1}{Z_q}\frac{-i\gamma^0p_0-i\sum_i\gamma^ip_i+m_r}
{p_0^2+\sum_ip_i^2+m_r^2},
\label{eq:matrix_trans}
\end{eqnarray}
in the sense that both expressions should have the same residue at the
heavy quark pole.

We begin by examining the numerator of the left-hand side of
Eq.~\ref{eq:matrix_trans} and substitute $p_0 \equiv i \tilde p_0$
\begin{equation}
\gamma^0 \sinh(\tilde p_0 a) -i \zeta\vec\gamma\cdot\vec p a
           + m_0 a +r_t\Bigl(1-\cosh(\tilde p_0 a)\Bigr) + \frac{r_s}{2}(\vec p a)^2.
\label{eq:spinor_matrix1}
\end{equation}
Two steps are needed to put this equation in a convenient form.  First we
replace the coefficient $\sinh(\tilde p_0 a)$ multiplying the $\gamma^0$ in
Eq.~\ref{eq:spinor_matrix1} by an expression closer to the continuum value:
\begin{equation}
\sinh(\tilde p_0 a) \approx \tilde p_0 a\frac{\sinh(m_r a)}{m_r a}.
\label{eq:approx_1}
\end{equation}
This approximation can be justified by using Eqs.~\ref{eq:pole_nzp2} and
\ref{eq:pole_nzp4} obeyed by $\tilde p_0$ to write:
\begin{equation}
\frac{\sinh(\tilde p_0 a)}{\tilde p_0} =
  \frac{\sinh(m_r a)}{m_r a}
  \left[\frac{1+\frac{(\vec p a)^2}{m_r a}\frac{\cosh(m_r a)}{sinh(m_r a)}}
             {1 + \frac{(\vec p a)^2}{(m_r a)^2} } \right]^{1/2}.
\end{equation}
We can then evaluate the difference between the contents of the square
bracket in this equation and one:
\begin{equation}
\Bigl[\ldots\Bigr]-1 = \frac{(\vec p a)^2}{m_r a} \left(
      \frac{\frac{\cosh(m_r a)}{\sinh(m_r a)} - \frac{1}{m_r a}}
           {1 + \frac{(\vec p a)^2}{(m_r a)^2}}\right) \le (\vec p a)^2.
\end{equation}
Here the final inequality, showing that this difference can be neglected,
follows from the relation $x\coth(x) \le (1+x+x^2)/(1+x)$.

The second relation that we need approximates:
\begin{eqnarray}
\cosh(\tilde p_0 a) &=& \cosh(m_r a)\left[1 + \frac{(\vec p a)^2}{m_r a}
                            \frac{\sinh(m_r a)}{\cosh(m_r a)}\right]^{1/2}
  \nonumber \\
&\approx& \cosh(m_r a)
                      +\frac{(\vec p a)^2}{2 m_r a}\sinh(m_r a).
\label{eq:approx_2}
\end{eqnarray}
Here the equality follows directly from Eq.~\ref{eq:pole_nzp2} while
the inequality requires the neglect of a term of order $(\vec p a)^4$
whose coefficient can be shown to be bounded through use of the relation
$\tanh(x) \le x$.

Next we substitute Eqs.~\ref{eq:approx_1} and \ref{eq:approx_2} into
Eq.~\ref{eq:spinor_matrix1} writing the numerator of the propagator as:
\begin{equation}
\gamma^0 \tilde p_0 a\frac{\sinh(m_r a)}{m_r a} -i \zeta\vec\gamma\cdot\vec p a
   + \sinh(m_r a) + \frac{1}{2}(\vec p a)^2\Bigl(r_s - r_t \frac{\sinh(m_r a)}{m_r a}\Bigr).
\label{eq:spinor_matrix2}
\end{equation}
We must now find matrices $U_L$ and $U_R$ which will transform this
expression into the desired continuum form.  Thus, we must make the
coefficient of $\vec \gamma \cdot \vec p$ agree with that of
$\gamma^0 p^0$ and remove the $\vec p\;^2$ term.  This can be accomplished
by matrices of the form
\begin{eqnarray}
U_L &=& U_R = (1 +i \delta \vec \gamma\cdot\vec p a) \quad \mbox{where}
\label{eq:spin_trans1} \\
\delta &=& \frac{\zeta}{2\sinh(m_r a)} -\frac{1}{2m_r a}
\label{eq:spin_trans2}
\end{eqnarray}
It is easy to see that when these transformations act on the $\sinh(m_r a)$
term in Eq.~\ref{eq:spinor_matrix2}, a term is generated which precisely
replaces $-i \zeta\vec\gamma\cdot\vec p a$ with the desired expression:
\newline
$-i\vec\gamma\cdot\vec p \sinh(m_r a)/m_r$.

However, the elimination of the $(\vec p a)^2$ term appearing in
Eq.~\ref{eq:spinor_matrix2} is less direct.  For this term the effect of
our transformation generates a $(\vec p a)^2$ contribution which is only
approximately zero:
\begin{equation}
(\vec p a)^2 \Bigl\{2\zeta(\frac{\zeta}{2\sinh(m_r a)} -\frac{1}{2m_r a})
   + \frac{1}{2}(r_s - r_t\frac{\sinh(m_r a)}{m_r a})\Bigr\}
             \approx 0.
\label{eq:p2_term}
\end{equation}
To neglect the expression in Eq.~\ref{eq:p2_term} we must make two observations.
First, we recognize that when expanded in a power series in $m_r a$ the
expression in curly brackets in Eq.~\ref{eq:p2_term} begins at order
$(m_r a)^1$ when $\zeta$ is determined by Eq.~\ref{eq:r_s-zeta_relation}.
This implies that for small $m_r a$, this unwanted $(\vec p a)^2$ term has
the size $(\vec p a)^2m_r a$ and is therefore $O(\vec p a)^2$ relative to
the mass term, $m_r a$.  Second, as $m_r a$ increases this expression grows
no faster than the other $\sinh(m_r a)$ factors in Eq.~\ref{eq:spinor_matrix2}.
Thus, the unphysical $(\vec p a )^2$ term is actually of order $(\vec p a )^2$
relative to the continuum terms in the Dirac propagator for all values of
$m_r a$.

The fact that a single choice of the transformation parameter $\delta$
is sufficient to both replace the coefficient of $\vec \gamma \cdot \vec p$
by its proper value and to remove the $(\vec p a)^2$ term is a result
of the relationship in Eq.~\ref{eq:r_s-zeta_relation} between $\zeta$ and
$r_s$, derived previously to insure the correct dispersion relation.

\subsection{Quark-gluon vertex}

We will now determine the parameters $c_B$ and $c_E$
by computing the quark-gluon vertex after the transformation of
Eq.~\ref{eq:spin_trans1} has been applied to the initial and final
spinors.  In particular, $c_E$ and $c_B$ should be chosen so that
the tree-level lattice vertex agrees with the corresponding continuum
expression, with errors no larger than $(\vec p' a)^2$,
$(\vec p a)^2$ and $\vec p'\cdot \vec p a^2$.  There should be no
contribution of order $(m_r a)^n$, $|\vec p\,' a|(m_r a)^n$ or
$|\vec p a|(m_r a)^n$ for all values of $n$.

Following the conventions listed in Appendix~\ref{ap:conventions},
we can determine the quark-gluon vertex $\Lambda_\mu(p',p)$ and
then  impose the on-shell conditions:
\begin{eqnarray}
\overline{u}(\vec p\,')\Lambda_\mu(p',p)u(\vec p)
                    = Z_q \overline{u}(\vec p\,')\gamma_\mu u(\vec p).
\end{eqnarray}
Here all quantities are evaluated following our Euclidean
space conventions with the exception of the time components
of the on-shell fermion momenta
$p'_0 = i\tilde p\,'_0 = i\sqrt{(\vec p\,')^2+m_r^2}$ and $p_0
= i\tilde p_0 = i\sqrt{(\vec p)^2+m_r^2}$.

The tree-level lattice vertex matrices $\Lambda_\mu(p',p)$ can be
derived from the lattice action of Eq.~\ref{eq:action_lat} and written
without approximation as:
\begin{eqnarray}
\Lambda^k(p',p) &=& \gamma^k\zeta \cos\left[(p_k'+p_k)a/2\right]
                    -i r_s\sin\left[(p_k'+p_k)a/2\right]
\nonumber \\
                &&+\frac{c_B}{2}\sum_j\sigma_{kj}\cos\left[(p'_k-p_k)a/2\right]
                                       \sin\left[(p'_j-p_j)a\right]
\nonumber \\
                &&+i\frac{c_E}{2}\sigma_{k0}\cos\left[(p'_k - p_k)a\right]
                                 \sinh\left[(\tilde p_0' - \tilde p_0)a\right]
\label{eq:vert_space_relation}
\\
\Lambda^0(p',p) &=& \gamma^0 \cosh\left[(\tilde p_0'+\tilde p_0)a/2\right]
                        + r_t\sinh\left[(\tilde p_0'+\tilde p_0)a/2\right]
\nonumber \\
                &&+\frac{c_E}{2}\sum_j\sigma_{0j}\cosh\left[(p'_0 - p_0)a/2\right]
                                 \sin\left[(p_j' - p_j)a\right].
\label{eq:vert_time_relation}
\end{eqnarray}

\subsubsection{Spatial component of the quark-gluon vertex}

We first examine the spatial quark gluon vertex $\Lambda^k$ transformed by
the spinor matrices $U_L(\vec p\,')$ and  $U_R(\vec p)$:

\begin{eqnarray}
[\Lambda_k(p',p)]_T &=& U_L(\vec p\,')^\dagger\Lambda_k(p',p)U_R(\vec p)^\dagger \\
                    &=& \zeta\gamma_k -i(\frac{r_s}{2}+\delta\zeta)(p_k+p'_k)a
  +(\frac{c_B}{2}+\delta\zeta)\sum_j\sigma_{kj}(p'_j-p_j)a
\nonumber \\
&&+i\frac{c_E}{2}\sigma_{k0}\sinh[(\tilde p'_0 -\tilde p_0) a]
\label{eq:vert_space_af_trans}
\end{eqnarray}
where the subscript $T$ indicates that we have applied the spinor
transformations $U_L(\vec p\,')$ and  $U_R(\vec p)$.  In addition, some terms of
relative order $(\vec p\,' a)^2$ and $(\vec p a)^2$have been neglected.

The expression in Eq.~\ref{eq:vert_space_af_trans} can be simplified if
we recognize that this matrix is to be evaluated between the spinors
$\overline u(\vec p\,')$ and $u(\vec p)$ so that we can use the relevant
Dirac equation:
\begin{eqnarray}
          \left(\gamma^0\tilde p_0 -i\vec\gamma\cdot\vec p  - m_r \right)u(\vec p) &=&0 \\
\overline u'(\vec p\,')
         \left(\gamma^0\tilde p'_0 -i\vec\gamma\cdot\vec p\,'  - m_r \right)       &=&0.
\end{eqnarray}
These two equations can be multiplied by $\gamma^k$ on the left and
right respectively to derive an equation for $\sigma_{kj}(p_j'-p_j)$:
\begin{equation}
\sigma_{kj}(p_j'-p_j) = 2m_r \gamma^k  + i(p_k'+p_k)
                       -i\sigma_{k0}(\tilde p_0'-\tilde p_0)
\label{eq:gordon_id_cont}
\end{equation}

Substituting the relation above for the $\sigma_{kj}$ term in
Eq.~\ref{eq:vert_space_af_trans}, we find
\begin{eqnarray}
[\Lambda_k(p',p)]_T &=&
\gamma^k\left(\zeta + m_r a(c_B + 2\delta\zeta)\right)\nonumber\\
&&+i\frac{c_B-r_s}{2}(p'_k+p_k)a\nonumber\\
&&+i\sigma_{k0}\left(\frac{c_E}{2}\sinh[(\tilde p'_0-\tilde p_0)a]
  -(\delta\zeta+\frac{c_B}{2})(\tilde p'_0-\tilde p_0)a\right)
\label{eq:vert_space_last}
\end{eqnarray}

The matrix $\Lambda^k(p',p)$ will reduce to the desired continuum
quantity $Z_q\gamma^k$ provided the following conditions are obeyed:
\begin{eqnarray}
&&c_B=r_s\label{eq:c_B_cond1}\\
&&\zeta + m_r a(c_B+2\delta\zeta)=Z_q \label{eq:c_B_cond2}\\
&&\overline u'(\vec p\,')\sigma_{k0}u(\vec p)
    \left(\frac{c_E}{2}\sinh[(\tilde p'_0-\tilde p_0)a]
    -(\delta\zeta+\frac{c_B}{2})(\tilde p'_0-\tilde p_0)a\right)
      = O(\vec{p} a)^2 Z_q \label{eq:c_B_E_cond}
\end{eqnarray}
We will treat the first of these conditions, Eq.~\ref{eq:c_B_cond1}, as
determining the quantity $c_B$.  The second equation,
Eq.~\ref{eq:c_B_cond2}, is then automatically obeyed as can be seen
by using $Z_q$ as determined by Eq.~\ref{eq:Z_q}, and substituting the
expressions given for $\delta$ and $\zeta$ in Eqs.~\ref{eq:spin_trans2}
and \ref{eq:r_s-zeta_relation} respectively.

Establishing the final condition, Eq.~\ref{eq:c_B_E_cond}, requires
a little more effort since it is not exact and must hold for the full
range of a variety of values of $r_s$ and $r_t$.  First we demonstrate
the argument of the difference $(\tilde p'_0 - \tilde p_0)a$ is small
so that an expansion of the $\sinh(x)$ function is justified.  We consider
the square:
\begin{eqnarray}
(\tilde p'_0 - \tilde p_0)^2a^2 &=& \left(\frac{(\tilde p'_0)^2 - (\tilde p_0)^2}
                                             {\tilde p'_0 + \tilde p_0}a\right)^2 \\
        &=& \frac{(\vec p\,')^2 - (\vec p)^2}{(\tilde p'_0 + \tilde p_0)^2}
              ((\vec p\,'a)^2 - (\vec p a)^2).
\label{eq:c_B_E_cond_prf_1}
\end{eqnarray}
Here the first factor on the right hand side of Eq.~\ref{eq:c_B_E_cond_prf_1}
is bounded for all values of $\vec p\,'$ and $\vec p$, while the second factor
is $O(\vec p a)^2$.  This justifies keeping only the first term in an expansion
of the $\sinh(x)$ function and replacing the third condition by
\begin{equation}
\overline u'(\vec p\,')\sigma_{k0} u(\vec p) \left(\frac{c_E-c_B}{2}
  -\delta\zeta\right)(\tilde p'_0-\tilde p_0)a
      = O(\vec{p} a)^2 Z_q
\label{eq:c_B_E_cond_prf_2}
\end{equation}

Next we make the choice $c_E = c_B$, multiply and divide the left
hand side of Eq.~\ref{eq:c_B_E_cond_prf_2} by $m_r a$ and divide by $Z_q$,
writing the resulting condition as:
\begin{equation}
\frac{m_r \overline u'(\vec p\,')\sigma_{k0} u(\vec p)}{\tilde p'_0+\tilde p_0}
            \left((\vec p\,'a)^2 - (\vec p a)^2\right)
            \frac{\delta\zeta}{m_raZ_q}
      = O(\vec{p} a)^2.
\label{eq:c_B_E_cond_prf_3}
\end{equation}
The left-most ratio in Eq.~\ref{eq:c_B_E_cond_prf_3} is a kinematic function,
which is bounded for all values of $m_r$.  The central factor provides the
desired $O(\vec pa)^2$ suppression.  We need to show that the final factor,
$\delta\zeta/(m_r a Z_q)$ is bounded for all $m_r a$.  To do this we must
require that for small $m_r a$, $r_t -r_s \propto m_r a$.  Without this
requirement, $\delta\zeta$ approaches a constant as $m_r a \rightarrow 0$ and
this factor diverges for small $m_r$ as $1/m_r a$.  Were we to choose
non-covariant values for $r_t$ and $r_s$ in the limit of small $m_r a$, then
the non-covariant choice $c_E \ne c_B$ would also be required.  For simplicity,
we make the choice $r_t=r_s$ for all values of $m_r$.  Under these
circumstances, it is easy to see by direct numerical evaluation that the
factor $\delta\zeta/(m_r a Z_q) \le 1/12$, its value at $m_r a=0$ for all
values of $r_s>0$ and $m_r a$.  Thus, condition~\ref{eq:c_B_E_cond_prf_3} is
also satisfied for the choice $c_E = c_B$ and the spatial components of the
quark gluon coupling agree with the expected continuum values to the
claimed accuracy.

\subsubsection{Temporal component of the quark-gluon vertex}

Finally we examine the time component of the quark-gluon vertex given in
Eq.~\ref{eq:vert_time_relation}.  As a first step we will simplify this
expression by recognizing that:
\begin{eqnarray}
\cosh[(\tilde p'_0 + \tilde p_0)a/2] &=& \cosh(m_r a) +O(\vec p a)^2 \\
\sinh[(\tilde p'_0 + \tilde p_0)a/2] &=& (\tilde p'_0 + \tilde p_0)
                          \frac{\sinh(m_r a)}{2 m_r a} +O(\vec p a)^2
\end{eqnarray}
and neglecting the $O(\vec p a)^2$ terms.  These two equations are
easy to derive from Eq.~\ref{eq:approx_1} using $\cosh(x) = \sqrt{\sinh^2(x)+1}$,
the formula for the hyperbolic sine of the sum of two angles and the
inequality $\sinh(x) \le x \cosh(x)$.  With these simplifications
$\Lambda^0$ becomes:
\begin{eqnarray}
\Lambda^0(p',p) &=& \gamma^0 \cosh(m_r a)
                     + r_t a (\tilde p'_0 + \tilde p_0)\frac{\sinh(m_r a)}{2 m_ra}
\nonumber \\
                   &&  + \frac{r_s}{2}\sum_j\sigma_{0j}(p'_j-p_j)a
\end{eqnarray}
where we have replaced $c_E$ by the value determined earlier, $c_E=c_B=r_s$.

Next, the spinor transformations $U_L(p')$ and $U_R(p)$ are made yielding
\begin{eqnarray}
\Lambda^0(p',p)_T  &=& U_L(p')^\dagger \Lambda^0(p',p) U_R(p)^\dagger \\
              &=& \gamma^0 \cosh(m_r a)
                     + r_t a (\tilde p'_0 + \tilde p_0)\frac{\sinh(m_r a)}{2 m_r a}
\nonumber \\
              &&  + \left(\frac{r_s}{2}+\delta\cosh(m_r a) \right)\sum_j\sigma_{0j}(p'_j-p_j)a.
\label{eq:c_B_E_cond_prf_4}
\end{eqnarray}
Here we have neglected the term:
\begin{equation}
\overline u'(\vec p\,')\gamma^j (p'_j+p_j)a u(p) \; r_t a(\tilde p'_0 + \tilde p_0)
                             \frac{\sinh(m_r a)}{2 m_r a}
\label{eq:c_B_E_cond_prf_5}
\end{equation}
because, as in the case of Eq.~\ref{eq:c_B_E_cond_prf_3}, the spinor structure
mixes upper and lower spinor components implying that this expression of order
$(\vec p a)^2$ and therefore negligible.

As a final step we use the time-component equivalent of Eq.~\ref{eq:gordon_id_cont}
multiplied by $r_t \sinh(m_r a)/(2 m_r a)$,
\begin{equation}
\frac{r_t \sinh(m_r a)}{2 m_r a} a (\tilde p'_0 + \tilde p_0)
                  = \frac{r_t \sinh(m_r a)}{2 m_r a}\Bigl\{2m_r a \gamma^0 -
                                    \sum_j\sigma_{0j}(p'_j-p_j)a\Bigr\},
\label{eq:c_B_E_cond_prf_6}
\end{equation}
to eliminate the $\tilde p'_0 + \tilde p_0$ term from Eq.~\ref{eq:c_B_E_cond_prf_4}.
The resulting expression is
\begin{eqnarray}
\Lambda^0(p',p)_T  &=& \gamma^0\Bigr\{\cosh(m_r a) + r_t \sinh(m_r a)\Bigr\}
                         \nonumber \\
            &&  + \Bigl\{\frac{r_s}{2}+\delta\cosh(m_r a)
           -r_t\frac{r_t \sinh(m_r a)}{2 m_r a}  \Bigr\}\sum_j\sigma_{0j}(p'_j-p_j)a.
\label{eq:c_B_E_cond_prf_7}
\end{eqnarray}
The first term in this equation is precisely the desired matrix $\gamma^0 Z_q$
while the second can be shown to be of order $(\vec p a)^2 Z_q$ using
the same style of argument that permitted us to neglect the similar term
in Eq.~\ref{eq:c_B_E_cond_prf_3} and the expression~\ref{eq:c_B_E_cond_prf_5}.

In conclusion, we have verified at tree level that only three, mass-dependent
parameters, $m_0$, $\zeta$ and $c_P=c_B=c_E$, are needed to realize
a heavy quark action that is accurate through order $|\vec{p}|a$
and to arbitrary order in $m_r a$.  We have the freedom to choose
$r_s$ and $r_t$ as is convenient but must require that as $m_r a$
approaches zero, $r_s \rightarrow r_t$.  The on-shell quark propagator
and quark-gluon vertex take their continuum form after a simple
$4 \times 4$ transformation is performed on the two external spinors.

\section{conclusion}
\label{sec:conclusion}

It is presently impractical to study charm or bottom physics
on a sufficiently fine lattice to control discretization errors
of order $ma$.  However, as established in Refs.~\cite{El-Khadra:1997mp}
and \cite{Aoki:2001ra} such errors can be avoided even when $ma \ge 1$
by using an improved heavy quark action.  Such an action will
accurately describe heavy quark states which are at rest
or have small spatial momenta and, as the quark mass is made
lighter or the lattice spacing finer, will smoothly approach
the usual $O(a)$-improved fermion action of Sheikholeslami and
Wohlert~\cite{Sheikholeslami:1985ij}.  Here we are referring
to this improved action as the ``relativistic heavy quark'' action
because of this smooth connection with relativistic fermions
as $ma \rightarrow 0$ and to distinguish it from the
non-relativistic and static approximations which do not have
this property.

By carrying out a systematic expansion in powers of $a$ but
working to all orders in the product $ma$, we have established
that only three parameters, $m_0$, $\zeta$ and $c_P$, need to
be tuned to remove all discretization errors of order
$\Lambda_{\rm QCD}$.  It is interesting to point out that the
possible overestimate of the number of relevant parameters in
the Fermilab and Tsukuba results was actually suggested to us by
the numerical work described in the companion paper~\cite{Lin:2006}.

In that paper we attempt to determine the relativistic heavy
quark parameters by a process of step scaling, beginning with
a very fine lattice where a direct use of the domain wall fermion
formulation gives accurate results.  We initially attempted to
determine the four parameters, $m_0$, $\zeta$, $c_B$ and $c_E$, that
could be used on at $16^3 \times 32$, $1/a=3.6$~GeV lattice to
reproduce the heavy-heavy and heavy-light spectra given by a
domain wall fermion calculation on a $24^3 \times 48$, $1/a=5.4$~GeV
lattice.  To our dismay, this was not a solvable problem, at
least with masses measured on the one percent level.  We found
a one-dimensional subspace in this four-dimension parameter
space along which all of the seven, finite-volume masses that
we computed did not change.  This surprising numerical result
lead us to study more closely the underpinnings for the
relativistic heavy quark formalism and to the 3-parameter result
presented here.

As is explained in detail in the companion paper, the problem
of determining three parameters by step-scaling is numerically
very stable and determining these heavy quark parameters to a few
percent is not difficult.  Although this first exploratory
numerical work is done within the quenched approximation, as
discussed in Ref.~\cite{Lin:2006}, we believe that similar results
will be possible in full QCD.  Thus, this approach to heavy
quark physics, especially in the charm region where only 2 or 3
step-scaling steps are needed, may provide a first-principles
approach with no reliance on perturbation theory.  The three
parameters needed in the heavy quark action as well as those
required for improved operators can be determined
non-perturbatively by this step-scaling approach, with the
three in the action requiring the most effort.

Further, as available resources increase, one can work at
increasingly fine lattice spacing, minimizing the higher order
errors that have not been explicitly removed.  No change in
formalism is needed.  However, extrapolation to the continuum
limit is in general not possible with this approach.  For
example, the $O(\Lambda_{\rm QCD}a)^2$ terms neglected in the
treatment above are expected to enter with coefficients
which are themselves functions of $ma$.  Thus, a simple $a^2$
behavior for small $a$ will be seen only in the limit
$ma \approx 0$, a region in which the improvements we have
discussed are not needed.

We would like Sinya Aoki, Peter Boyle, Changhoan Kim, Yoshinobu
Kuramashi, Chris Sachradja and our colleagues in the RBC
collaboration for helpful discussions and suggestions.  This
work was supported in part by DOE Grant DE-FG02-92ER40699.

\appendix
\section{Conventions}
\label{ap:conventions}
As described in Sec.~\ref{sec:analysis}, we use hermitian Dirac
gamma matrices, appropriate for Euclidean lattice QCD calculations,
which satisfy $\{\gamma^\mu, \gamma^\nu\} = 2 \delta^{\mu,\nu}$.
In addition, we use the matrices
$\sigma_{\mu,\nu} = \frac{i}{2}[\gamma^\mu,\gamma^\nu]$.  Since the
the methods discussed here are to be applied in the approximate
rest system of the heavy quark, when an explicit choice for the
gamma matrices is needed, we adopt conventions where $\gamma^0$
is diagonal.  Specifically, in terms of standard
$2 \times 2$ blocks, we use:
\begin{equation}
\gamma^0=\left(
\begin{array}{cc}
I  &0  \\
0  &-I
\end{array} \right)
\quad
\gamma^i=\left(
\begin{array}{cc}
0  &-i\sigma^i  \\
i\sigma^i  &0
\end{array} \right)
\end{equation}
where $I$ is the $2 \times 2$ identity matrix and the $\sigma^i$ are
the standard Pauli matrices.

With this choice of gamma matrices, a spinor solution of the
Dirac equation, describing an on-shell particle with 3-momentum $\vec p$
and energy $\tilde p_0 = \sqrt{\vec p^2 + m_r^2}$, can be written
\begin{eqnarray}
u_s(\vec p) &=& \left(
\begin{array}{c}
\chi_s \\
\frac{\vec\sigma \cdot\vec p}{\tilde p_0 +m}\chi_s
\end{array} \right)
\end{eqnarray}
where $\chi_s$ is a two-component column vector describing the two
possible spin-1/2 states labeled by $s = \pm 1/2$.

We determine the tree-level quark-vertex from the lattice
action of Eq.~\ref{eq:action_lat} by replacing the link variables
$U_\mu(n)$ by the combination $1 -igt^a A_\mu^a(n+\hat e_\mu/2)$,
and writing the external field $A_\mu^a(n+\hat e_\mu /2)$ in terms of
the Euclidean Fourier transform:
\begin{equation}
A_\mu^a(n+\hat e_\mu /2)
    = (\frac{a}{2\pi})^2\prod_{\nu=0}^{3} \int_{\pi/a}^{-\pi/a} d q_\nu
                       e^{iq\cdot(n+ \hat e_\mu/2)a}{\cal A}_\mu^a(q).
\end{equation}
Here the matrices $t^a$ are hermitian generators of the gauge group.
The matrix coefficient of the amplitude ${\cal A}_\mu(p'-p)$ appearing
in the tree-level evaluation of the action is identified as
$-ig  \overline u'(\vec p\,') t^a \Lambda_\mu(p',p)u(\vec p)$.

Note with the relationship between $U_\mu(n)$ and $A_\mu^a(n+\hat e_\mu /2)$
adopted above, our covariant derivatives become:
\begin{eqnarray}
D_\mu \psi &=& (\partial_\mu -i g t^a A^a_\mu)\psi
\label{eq:cov_derivative_1} \\
\overline{\psi}\overleftarrow{D}_\mu \psi
           &=& \overline{\psi}(\overleftarrow{\partial}_\mu +i g t^a A^a_\mu)
\label{eq:cov_derivative_2}
\end{eqnarray}

\bibliography{06th_rhq}

\begin{thebibliography}{6}
\expandafter\ifx\csname natexlab\endcsname\relax\def\natexlab#1{#1}\fi
\expandafter\ifx\csname bibnamefont\endcsname\relax
  \def\bibnamefont#1{#1}\fi
\expandafter\ifx\csname bibfnamefont\endcsname\relax
  \def\bibfnamefont#1{#1}\fi
\expandafter\ifx\csname citenamefont\endcsname\relax
  \def\citenamefont#1{#1}\fi
\expandafter\ifx\csname url\endcsname\relax
  \def\url#1{\texttt{#1}}\fi
\expandafter\ifx\csname urlprefix\endcsname\relax\def\urlprefix{URL }\fi
\providecommand{\bibinfo}[2]{#2}
\providecommand{\eprint}[2][]{\url{#2}}

\bibitem[{\citenamefont{El-Khadra et~al.}(1997)\citenamefont{El-Khadra,
  Kronfeld, and Mackenzie}}]{El-Khadra:1997mp}
\bibinfo{author}{\bibfnamefont{A.~X.} \bibnamefont{El-Khadra}},
  \bibinfo{author}{\bibfnamefont{A.~S.} \bibnamefont{Kronfeld}},
  \bibnamefont{and} \bibinfo{author}{\bibfnamefont{P.~B.}
  \bibnamefont{Mackenzie}}, \bibinfo{journal}{Phys. Rev.}
  \textbf{\bibinfo{volume}{D55}}, \bibinfo{pages}{3933} (\bibinfo{year}{1997}),
  \eprint{hep-lat/9604004}.

\bibitem[{\citenamefont{Aoki et~al.}(2003)\citenamefont{Aoki, Kuramashi, and
  Tominaga}}]{Aoki:2001ra}
\bibinfo{author}{\bibfnamefont{S.}~\bibnamefont{Aoki}},
  \bibinfo{author}{\bibfnamefont{Y.}~\bibnamefont{Kuramashi}},
  \bibnamefont{and} \bibinfo{author}{\bibfnamefont{S.-i.}
  \bibnamefont{Tominaga}}, \bibinfo{journal}{Prog. Theor. Phys.}
  \textbf{\bibinfo{volume}{109}}, \bibinfo{pages}{383} (\bibinfo{year}{2003}),
  \eprint{hep-lat/0107009}.

\bibitem[{\citenamefont{Kronfeld}(2004)}]{Kronfeld:2003sd}
\bibinfo{author}{\bibfnamefont{A.~S.} \bibnamefont{Kronfeld}},
  \bibinfo{journal}{Nucl. Phys. Proc. Suppl.} \textbf{\bibinfo{volume}{129}},
  \bibinfo{pages}{46} (\bibinfo{year}{2004}), \eprint{hep-lat/0310063}.

\bibitem[{\citenamefont{Lin and Christ}(2006)}]{Lin:2006}
\bibinfo{author}{\bibfnamefont{H.-W.} \bibnamefont{Lin}} \bibnamefont{and}
  \bibinfo{author}{\bibfnamefont{N.~H.} \bibnamefont{Christ}}
  (\bibinfo{year}{2006}), \eprint{hep-lat/0608005}

\bibitem[{\citenamefont{Luscher}(1977)}]{Luscher:1976ms}
\bibinfo{author}{\bibfnamefont{M.}~\bibnamefont{Luscher}},
  \bibinfo{journal}{Commun. Math. Phys.} \textbf{\bibinfo{volume}{54}},
  \bibinfo{pages}{283} (\bibinfo{year}{1977}).

\bibitem[{\citenamefont{Sheikholeslami and
  Wohlert}(1985)}]{Sheikholeslami:1985ij}
\bibinfo{author}{\bibfnamefont{B.}~\bibnamefont{Sheikholeslami}}
  \bibnamefont{and} \bibinfo{author}{\bibfnamefont{R.}~\bibnamefont{Wohlert}},
  \bibinfo{journal}{Nucl. Phys.} \textbf{\bibinfo{volume}{B259}},
  \bibinfo{pages}{572} (\bibinfo{year}{1985}).

\end{thebibliography}

\begin{table}
\centering
\caption{Enumeration of the transformation coefficients that have been
chosen to remove redundant terms in ${\cal L}_{\rm eff}$ and the result
achieved.}
\label{tab:improvement}
\begin{tabular}{llc}
\hline\hline
Coefficient                 & Term in ${\cal L}_{\rm eff}$ that was improved & Order\\
\hline
$r_{N_0+1}^{0,1}$
                            & Coef. of $\gamma^0D^0$ set to 1               & $1/a$ \\
$s_{N_0+1}^{0,1}$
                            & Coefs. $\propto (aD^0)^2$ removed from
                                     the mass term                          & $1/a$ \\
$r_{N_0+1}^{1,1}$
                            & Coef. of $\vec D^2$ set to 0                  & $a$   \\
$s_{N_0}^{1,1}$
                            & Coefs. $\propto (aD^0)^2$ removed from
                                     $\gamma \vec D$ term                   & $a^0$ \\
$r_{N_0+1}^{2,1}$
                            & Coef. of $a\{\vec D^2,a \gamma^0 D^0\}$
                                  set to 0                                  & $a$   \\
$s_{N_0}^{2,2}$
                            & Coefs. $\propto (aD^0)^2$ removed from
                                       $[\gamma^i,\gamma^j][D^i,D^j]$ term  & $a$   \\
$r_{N_0+1}^{2,2}$
                            & Coef. of
                $a\{[\gamma^i,\gamma^j][D^i,D^j],a\gamma^0 D^0\}$ set to 0  & $a$   \\
$s_{N_0}^{2,3}$
                            & Coefs. $\propto (aD^0)^2$ removed from
                                       $[\gamma^i,\gamma^0][D^i,D^0]$ term  & $a$   \\
$r_{N_0}^{2,3}$
                            & Equate $[\gamma^i,\gamma^0][D^i,D^0]$ coef. to that
                                    of $[\gamma^i,\gamma^j][D^i,D^j]$       & $a$   \\
\hline

\end{tabular}
\end{table}

\clearpage

%%%%%%%%%%%%%%%%%%%%%%%%%%%%%%%%%%%Figures%%%%%%%%%%%%%%%%%%%%%%%%%%%%%%%%%%%%%%%%

%Z-factor diagram.

\end{document}